\documentclass[floatfix,aps,preprint,showpacs,amsmath,amssymb]{revtex4}
\usepackage[latin1]{inputenc}

\usepackage[dvips]{graphicx}

\usepackage{dcolumn}
\usepackage{bm}
\usepackage{dsfont}
\usepackage{color}
\usepackage{ulem} 
\usepackage{url}
\usepackage[colorlinks=true ,citecolor=blue]{hyperref}
\usepackage{amsmath,amssymb}
\newcommand{\sgn}{\text{sgn}}

\newcommand{\Pf}{\text{Pf}}

\begin{document}

\title{ Introduction to Topological Phases and Electronic Interactions in (2+1) Dimensions.}

\author{Leandro O. Nascimento$^{1,2}$}
\affiliation{$^1$International Institute of Physics, Campus Universitário-Lagoa Nova-59078-970-CP: 1613 - Natal/RN-Brazil \\
$^2$Faculdade de Ci\^encias Naturais, Universidade Federal do Par\'a, C.P. 68800-000, Breves, PA,  Brazil}

\date{\today}

\begin{abstract}

A brief introduction to topological phases is provided, considering several two-band Hamiltonians in one- and two-dimensions. Relevant concepts of the topological insulator theory, such as: Berry phase; Chern number; and the quantum adiabatic theorem, are reviewed in a basic framework, which is meant to be accessible to non-specialists. We discuss the Kitaev chain, SSH, and BHZ models. The role of the electromagnetic interaction in the topological insulator theory is addressed in the light of the Pseudo quantum electrodynamics (PQED). The well known parity anomaly for massless Dirac particle is reviewed in terms of the Chern number. Within the continuum limit, a half-quantized Hall conductivity is obtained. Thereafter, by considering the lattice regularization of the Dirac theory, we show how one may obtain the well known quantum Hall conductivity for a single Dirac cone. The renormalization of the electron energy spectrum, for both small and large coupling regime, is derived. In particular, it is shown that massless Dirac particles may, only in the strong correlated limit, break either chiral or parity symmetries. For graphene, this implies the generation of Landau-like energy levels and the quantum valley Hall effect.

\begin{center}
Subject Areas: Condensed Matter Physics, Graphene, Field Theory Methods
\end{center}

\end{abstract}


\maketitle

\section{\textbf{I.\,Introduction}}

The experimental realization of two-dimensional materials, for instance: graphene \cite{grapexp}; silicene \cite{silexp}; phosphorene \cite{phos}; germanene \cite{germanene}; and transtion metal dichalcogenides \cite{TMD}, has allowed a lively interaction between the communities of condensed matter and quantum field theory physicists, leading to several propositions of new topological states of matter with applications that range from nanoelectronics to quantum computation \cite{Bernevig}. The fundamental link between the high- and low-energy physics is the existence either of the so-called Dirac cones or the massive Dirac theory in the low-energy description of the these crystals. The Dirac theory of the crystal appears around some special point in the first Brillouin zone, for instance, we may cite the K and K' valleys of the honeycomb lattice \cite{review}. Topological phases are described by topological order instead of spontaneous symmetry breaking.      

The discovery of topological order has been one of the most important advances in condensed matter physics. It has extended the well-developed band theory and open new areas, such as spintronics and valleytronics. For one dimensional systems, the topological order has been discussed either in the context of Majorana fermions (in Kitaev chain) \cite{Kitaev} or zero-energy modes of the SSH model \cite{SSH}. For two-dimensional systems, the topological order has been used to explain a novel of quantum transport effects, such as the QSHE \cite{QSHE}, QVHE \cite{prx}, and QAHE \cite{Haldane}. In these cases, the systems are called topological insulators, because they are insulators in the bulk of the material, but they are conductors at their edges. These edge states are zero-energy modes that cross the Fermi level. The topological insulators, for noninteracting systems, have been classified in a ten-fold table, which takes into account a set of discrete symmetries in order to classify each case. These symmetries are time-reversal symmetry, charge conjugation, and chiral symmetry. They are discrete symmetries, which explain why spontaneous symmetry breaking (it works only for continuous symmetries) does not explain these effects. Nevertheless, the role of electronic interactions is still to be understood in general grounds \cite{Bernevig}.

It is well known that the dynamical mass generation for massless Dirac particle breaks either chiral symmetry or parity symmetry \cite{Appel}. Since the mass term modifies the topological properties of the Dirac theory, it is important \textit{always} to investigate whether strong coupling constants may renormalize the electron energy spectrum. Recently, it has been shown that electronic interactions in graphene generate a spontaneous quantum valley Hall effect at large coupling constant and low temperatures. In this effect, there exist two-counter propagating currents at the edge of the sheet and the difference between them yields a Hall conductivity equal to the usual result in the presence of an external magnetic field, namely, $\sigma^0_{xy}=4(n+1/2)e^2/h$ \cite{prx}. The authors have considered the Pseudo quantum electrodynamics (PQED), which is the natural theory for describing electron-electron interaction in the plane \cite{marino}. The key aspect for obtaining this result was the application of the Schwinger-Dyson equations for the electron self-energy, which is a typical approach for strong correlated systems in quantum field theory.

Although the concepts of topological insulator theory is well known by the community of condensed matter physics, perhaps quantum-field-theory physicists   may be confused, in a first view to topological phases, with several different results and concepts. Since the quantum field theory has developed several powerful methods for strongly correlated systems, it is relevant to provide a self-consistent introduction to topological insulator theory, using a common language, yielding a better connection between Dirac-like models and two-dimensional materials. This is the main goal of the first part of this paper. In particular, we would like to clarify how electronic interactions may change the electronic spectrum, hence, potentially generating a topological phase as it has been obtained in Ref.~\cite{prx}. This is one goal of part B.

The outline of this paper is the following: In part A, we review the main concepts related to topological phases in one and two dimensions. In Section A.~I, we review the derivation of the Berry phase. In Section A.~II, we calculate the Berry phase for one- and two-dimensional Hamiltonian with two energy bands. Furthermore, we define the Majorana number. In Section A.~III, we derive the quantum adiabatic theorem. In Section A.~IV, we apply the quantum adiabatic theorem in order to show that the Hall conductivity is given by the sum of the Berry phase of the filled energy bands (below the Fermi level). In Section A.~V, we review the SSH and Kitaev models. In Section A.~VI, we discuss the parity anomaly in massless Dirac theory and the BHZ model. In part B, we discuss the role of electronic interactions for two-dimensional materials. In Section B.~I, we show that, at small coupling, there is no dynamical mass generation. In Section B.~II, we calculate both the longitudinal and Hall conductivity for massive Dirac fermions.

\section{Part A: Topological Phases.}

Here, we provide a short review about topological phases for two-band Hamiltonians in one- and two-dimensions. We focus on analytical calculations of the Berry phase for SSH, Dirac, and BHZ models. We show that the Kitaev chain has Majorana fermions. Furthermore, we derive the quantum adiabatic theorem and use it to show the relation between the Hall conductivity and Berry phase.

\section{A.~I-The Berry phase.}
\label{cap52} 

In general grounds, the Berry phase $\gamma$ is generated in the eigenstate $|\Psi \rangle$ of the Hamiltonian $H(t)$, when the parameter changes from $t$ to $t+T$ with $H(t+T)=H(t)$, where $T$ is some period. In this case, $|\Psi\rangle\rightarrow \exp(i\gamma)|\Psi\rangle$, after one cycle. The most well known systems with such periodicity are the crystals in solid state physics, where the Bloch Hamilonian $H(k)$ is invariant under translational symmetry, which implies $H(k)=H(k+2\pi/a)$, where $k$ is the momentum in the first Brillouin zone, and $a$ is the lattice parameter. This is a consequence of the boundary condition applied for the crystal, i.e, $|\Psi(R+L)\rangle=|\Psi(R)\rangle$, where $L=N a$ with $N\approx 10^{24}$ being the number of atoms and $R$ the position of the atoms.

It is possible to show that the Berry phase is a topological invariant, for instance, in one dimension is related to the winding number and in two dimensions to the Chern number. Hence, the Berry phase has been used  to classify either trivial phase (with zero Berry phase) or topological phase (with nonzero Berry phase). Due to its topological feature, the Berry phase is \textit{invariant} under a large class of continuous transformations. These transformations are called adiabatic transformations in which the parameter slowly varies in time, i.e, $dk/dt \ll 1$. For a gapped system, we have $k \Delta \ll 1$, where $\Delta$ is the gap between two eigenstates. Systems with a finite energy gap are called insulators. Hence, unless the system undergoes a topological phase transition that would change its Berry phase, they all belong to the same class with the same Berry phase. Here, we are interested in the class of systems which have nonzero Berry phase and are called topological insulators \cite{Bernevig}. Next, we derive the Berry phase.

We assume a system described by the Hamiltonian $H(\textbf{k}(t))$, which is dependent on the set of time-dependent parameters $\textbf{k}(t)$. The eigenstates are solutions of the time-dependent equation
\begin{equation}
i \hbar \frac{\partial}{\partial t} |\Psi(t)\rangle=H(t) |\Psi(t) \rangle,  \label{eqsch00x}
\end{equation}
where we have simplified the notation, using that $|\Psi(\textbf{k}(t))\rangle \equiv |\Psi(t) \rangle$. Next, we consider the time-independent solution $|n(0) \rangle$ that satisfies the equation
\begin{equation}
H(0)|n(0)\rangle=E_n |n(0) \rangle. \label{eqsch01}
\end{equation}
The time-dependent eigenstates are $|n(t) \rangle=U(t,0) |n(0) \rangle$, where $U(t,0)$ is the usual time-evolution operator. Because of $U^{\dagger} U=1$, they are a normalized basis in the Hilbert space. Therefore, we may expand the eigenstates $|\Psi(t) \rangle$ as the following  
\begin{equation}
|\Psi(t) \rangle=\sum_n | \psi_n(t) \rangle =\sum_n e^{-i \theta_n(t)/\hbar} a_n(t) |n(t) \rangle, \label{exppsi02}
\end{equation}
where the $\theta_n(t)$ are arbitrary phases for each $n$ state. We could use Eq.~(\ref{exppsi02}) in Eq.~(\ref{eqsch0x}) what would lead us to the quantum adiabatic theorem, which we shall discuss in Section A.~III. Here, we only use the simplest solution of this theorem that yields $a_n(t)=a_n(0)$. Without loss of generality, we choose $a_n(0)=\delta_{nl}$, which is a projection into the $l$-state. Using this condition in Eq.~(\ref{exppsi02}), we have
\begin{equation}
|\psi_l(t) \rangle= e^{-i \theta_l(t)/\hbar} |l(t) \rangle. \label{exppsi03}
\end{equation}

The eigenvalue equation reads
\begin{equation}
i \hbar \frac{\partial}{\partial t} |\psi_l(t)\rangle=H(t) |\psi_l(t) \rangle=E_l(t) |\psi_l(t) \rangle.  \label{eqsch0x}
\end{equation}

Using Eq.~(\ref{exppsi03}) in Eq.~(\ref{eqsch0x}), we obtain
\begin{equation}
\hbar \frac{d\theta_l(t)}{dt}|l(t)\rangle+i\hbar \frac{d|l(t)\rangle}{dt}=E_l(t)|l(t)\rangle. \label{thetaeq}
\end{equation}

Multiplying by $\langle l(t)|$ both sides of Eq.~(\ref{thetaeq}), we obtain the phase $\theta_l(t)$, given by
\begin{equation}
\theta_l(t)=\frac{1}{\hbar}\int_0^t E_l(t')dt'-\gamma_l(t), \label{totph}
\end{equation}
where
\begin{equation}
\gamma_l(t)=i \int_0^t \langle l(t')|\frac{d}{dt'}|l(t')\rangle dt' \label{bergen}
\end{equation}
is the so-called Berry phase. It is a contribution only dependent on the eigenstates of the theory, hence, it is also called a geometric phase. On the other hand, the first term in the right-hand side (rhs) of Eq.~(\ref{totph}) is the famous dynamical phase, dependent on the eigenvalues. 

From now on, we apply Eq.~(\ref{bergen}) for solid state physics, thereby, the relevant parameter is the momentum $\textbf{k}$ in the first Brillouin zone (B.Z). In this context, the Berry phase reads
\begin{equation}
\gamma_l=i \int_{\textbf{k}\,\in\, B.Z} \langle l(\textbf{k})|\nabla_\textbf{k} |l(\textbf{k})\rangle. d\textbf{k}=\oint d\textbf{k}.\textbf{A}(\textbf{k}), \label{berrpha}
\end{equation}
where 
\begin{equation}
\textbf{A}(\textbf{k})=-\textbf{Im} \langle l(\textbf{k})|\nabla_\textbf{k} |l(\textbf{k})\rangle \label{bervec}
\end{equation}
is called the Berry vector and the integral in Eq.~(\ref{berrpha}) is over the first Brillouin zone, which is a closed path in the k-space. Eq.~(\ref{berrpha}) may be converted into a surface integral, yielding
\begin{equation}
\gamma_l=\int (\nabla_\textbf{k}\times \textbf{A}). d^2\textbf{k}=\int \epsilon^{ijk}(\partial_{j}A_k) d^2k_i, \label{divs}
\end{equation}
where $\partial_j\equiv \partial_{kj}\equiv \partial/\partial k_j$. Using Eq.~(\ref{bervec}) in Eq.~(\ref{divs}), we have
\begin{equation}
\gamma_l=-\textbf{Im} \int_S d^2k_i \epsilon^{ijk}\langle \partial_j l(\textbf{k})| \partial_k l(\textbf{k})\rangle. \label{eqg0}
\end{equation}

Next, we introduce the identity $1=\sum_m |m(\textbf{k}) \rangle \langle m(\textbf{k})|$ in Eq.~(\ref{eqg0}). Neglecting the term $m=l$, which does not contribute to the Berry phase, we find
\begin{equation}
\gamma_l=-\textbf{Im}\int_Sd^2k_i \sum_{m\neq l}\epsilon^{ijk}\langle \partial_j l(\textbf{k})|m(\textbf{k})\rangle \langle m(\textbf{k})|\partial_k l(\textbf{k})\rangle. \label{eqg1}
\end{equation}

It is convenient to eliminate the derivatives of the eigenvectors in Eq.~(\ref{eqg1}). Note that  and $E_l \langle m|\partial_{\textbf{k}} l \rangle= \langle m|\partial_{\textbf{k}} H l \rangle=\langle m|(\partial_{\textbf{k}} H) |l \rangle+E_m \langle m|\partial_{\textbf{k}} l \rangle $. Therefore, we find a very useful identity given by
\begin{equation}
\langle m|\partial_{\textbf{k}} l \rangle=\frac{\langle m|(\partial_{\textbf{k}} H) |l \rangle}{E_l-E_m}. \label{propri}
\end{equation}
Using Eq.~(\ref{propri}) in Eq.~(\ref{eqg1}), we obtain the Berry phase, in its gauge-invariant equation, given by
\begin{equation}
\gamma_l=\int_S d^2\textbf{k}. \textbf{V}_l, \label{bermag}
\end{equation} 
where $\textbf{V}_l$ is similar to a magnetic field in the k-space and reads
\begin{equation}
\textbf{V}_l=-\textbf{Im}\sum_{m\neq l}\frac{\langle l(\textbf{k})|(\nabla_\textbf{k} H)|m(\textbf{k})\rangle\times \langle m(\textbf{k})|(\nabla_\textbf{k} H)|l(\textbf{k})\rangle}{(E_l(\textbf{k})-E_m(\textbf{k}))^2}.
\end{equation}

For a two-dimensional system, the surface is in the $k_x-k_y$ plane, then $d^2\textbf{k}=dk_x dk_y \hat k_z$ with $\hat k_z=\hat k_x\times\hat k_y$. Therefore, from Eq.~(\ref{eqg0}), we obtain that the Berry phase is written as
\begin{equation}
\gamma_l=-\textbf{Im}\int_S dk_x dk_y \Omega^l_{k_xk_y}, \label{berphcur}
\end{equation} 
where $\Omega^l_{k_xk_y}$ is the Berry curvature given by 
\begin{equation}
\Omega^l_{k_xk_y}=\langle \partial_{k_x} l(\textbf{k}) |\partial_{k_y}l(\textbf{k})\rangle -h.c. \label{bercur}
\end{equation}

From Eq.~(\ref{bercur}), we conclude that the Berry curvature is an imaginary function, therefore, according to Eq.~(\ref{berphcur}), the Berry phase is real, as expected. For one and two-dimensional systems the Berry phase may be calculated from Eq.~(\ref{berrpha}) and Eq.~(\ref{bermag}), respectively. 

\section{A.~II-Topological invariants for two-band systems in one and two dimensions.}
\label{cap52} 

In this Section, we calculate the Berry phase for one- and two-dimensional Hamiltonians. For the sake of simplicity, we shall discuss only two-band systems, which are the relevant cases to compare with the Dirac theory.

The most generic Hamiltonian for a two-band system is given by
\begin{equation}
H=\textbf{h}(\textbf{k}).\vec{\sigma}=\left(\begin{matrix} h_3& h_1+ih_2 \\ h_1-ih_2&-h_3 \end{matrix}\right), \label{h2}
\end{equation}
where $\textbf{h}=(h_1,h_2,h_3)$ is an arbitrary three-vector and $\vec{\sigma}=(\sigma_x,\sigma_y,\sigma_z)$ are the Pauli matrix. The two eigenvalues are $E_{\pm}=\pm |h|=\sqrt{h_1^2+h_2^2+h_3^2}$. The eigenvectors are such that $H|\pm \rangle =E_{\pm} |\pm \rangle$, $\langle +|+\rangle=1$, $\langle -|-\rangle=1$, $\langle +|-\rangle=0$, and $\langle -|+\rangle=0$. Therefore,
\begin{equation}
\left(\begin{matrix} h_3& h_1+ih_2 \\ h_1-ih_2&-h_3 \end{matrix}\right)\left(\begin{matrix} a \\ b \end{matrix}\right)=E_{\pm} \left(\begin{matrix} a \\ b \end{matrix}\right),
\end{equation}
and
\begin{equation}
(h_1-ih_2)a=(h_3+E_{\pm})b. \label{evcre}
\end{equation}

From Eq.~(\ref{evcre}) and using the normalization of the eigenvectors, we find
\begin{equation}
|\pm \rangle={\cal N}_{\pm}\left(\begin{matrix} h_3+E_{\pm} \\ h_1-ih_2 \end{matrix}\right), \label{evh2}
\end{equation}
where
\begin{equation}
{\cal N}_{\pm}=\frac{1}{\sqrt{2E_{\pm}(E_\pm+h_3)}}
\end{equation}
is the normalization constant. 

The wave function in Eq.~(\ref{evh2}) is in a particular choice of gauge. Indeed, we may find other solutions by performing $|\pm\rangle \rightarrow e^{i\varphi}|\pm\rangle $. It implies, according to Eq.~(\ref{bervec}), that the Berry vector changes to $\textbf{A}\rightarrow \textbf{A}-\nabla_\textbf{k} \varphi$. Assuming our eigenstates $|\pm \rangle$ are single-valued in the Brillouin zone, i.e, $|\pm \rangle (k=+\pi/a)=|\pm \rangle (k=-\pi/a)$, hence, we find that $\varphi(k=+\pi/a)-\varphi(k=-\pi/2)=2\pi n$ with $n$ integer. Nevertheless, note that we have obtained a gauge-invariant equation for the Berry phase in Eq.~(\ref{divs}), because $\nabla_k\times \textbf{A}$ is invariant under the gauge transform. 

\subsection{The One Dimensional Case}

For one dimensional systems, we shall discuss two topological invariants, one is the Berry phase for off-diagonal Hamiltonians ($h_3=0$), and the other is the Majorana number $M$ for a Hamiltonian with $h_3\neq 0$. Let us first discuss the Berry phase.

We consider $h_3=0$ in Eq.~(\ref{h2}) and Eq.~(\ref{evh2}). Thereby, the eigenstates are
\begin{equation}
|\pm \rangle=\frac{1}{\sqrt{2}}\left(\begin{matrix} \pm 1 \\ \hat h^*(k) \end{matrix}\right), \,\, \hat h(k)\equiv\frac{h(k)}{|h(k)|}. \label{1dev}
\end{equation}
 
The Berry phase is obtained from Eq.~(\ref{berrpha}) and, for the valence band $E=E_-$, is given by
\begin{eqnarray}
\gamma_l(E_-)&=&-\int_{B.Z} dk A(k)=-2 i\int_{B.Z} \langle -|\partial_k|-\rangle dk\nonumber\\
&=&-i \int_{B.Z} dk\,[\hat h^*(k)]^{-1}\partial_k \hat h^*(k), \label{ber1d}
\end{eqnarray}
where the extra factor $2$ is due to the two spins up and down. The minus sign is due to the general relation $\gamma_l(E_-)=-\gamma_l(E_+)$, which shows that the valence band has opposite sign in comparison with the conduction band. Using $dh^*=dk \partial h/\partial k$, Eq.~(\ref{ber1d}) becomes
\begin{equation}
\gamma_l(E_-)= \frac{1}{i}\oint_{C_h}\frac{d\hat h^*}{\hat h^*}=\frac{1}{i}\oint_{C_h}\frac{d h^*}{ h^*}, \label{ber1d2}
\end{equation}
where $C_h$ is a closed path in the $h$-space. This is because the integral in $k$ is over the first Brillouin zone with $h(k_x+2\pi,k_y+2\pi)=h(k_x,k_y)$. The topological invariant in Eq.~(\ref{ber1d2}) is conventionally called Zak phase instead of 
Berry phase in 1D. By using ultra-cold atoms in optical lattices, the authors in Ref.~\cite{zakphase} have measured the difference of the two Zak phases of the Rice-Mele model, which mimics polyacetylene depending on its parameters values. This model also has two topological phases. Although, the Zak phase is gauge-dependent, its difference is uniquely defined.

Eq.~(\ref{ber1d2}) has a geometrical interpretation. Indeed, let us consider the winding number $w$ of a closed curve $C$ in the complex plane around a point $a$. In complex analysis, this is given by
\begin{equation}
w=\frac{1}{2\pi i} \oint_{C} \frac{dh}{h-a_0}, \label{windCA}
\end{equation} 
where $h=h_x+ih_y=r\exp(i\theta)$ is a complex number. By comparison between Eq.~(\ref{ber1d2}) and Eq.~(\ref{windCA}), we conclude that: \textit{The winding number is the Berry phase divided by} 2$\pi$. We shall apply these results for the SSH model.

Next, let us consider $h_3 \neq 0$. In the basis ${\cal C}^\dagger_k\equiv (c_k \,\,\, c^\dagger_{-k})$, the Hamiltonian in Eq.~(\ref{h2}) reads $H={\cal C}^\dagger_k h(k) {\cal C}_k$. We define a new basis given by
\begin{equation}
a_k=c_k+c^\dagger_{-k}, \,\, i b_k=c_k-c^\dagger_{-k}. \label{majopt}
\end{equation}

After some calculation, one may show that the Hamiltonian in this basis is given by $H=1/2 A^\dagger B(k) A$, where $A^\dagger=(a_k\,\, b_k)$, $B(k)=\textbf{b}.\sigma$, and $\textbf{b}=(-h_2,-h_3,h_1)$. The Majorana number $M$ is defined as the sign of the product of the Pfaffian of $B(k)$ at $k=\pi$ and $k=0$, i.e, $M= \sgn\{\Pf [B(k=\pi)]\} \sgn\{\Pf [B(k=0)]\}$ \cite{Kitaev}. The Pfaffian of $B(k)$ is $\Pf[B(k)]=\pm \sqrt{\det B(k)}=\sgn(E_{\pm}) |E_{\pm}(k)|$. Therefore,
\begin{equation}
M=\sgn[h_3(k=\pi)]\sgn[h_3(k=0)]. \label{Majonumber}
\end{equation}

In Eq.~(\ref{Majonumber}) we have used that the sign of $h_3$ is the sign of the energy band. This is because the $z$-axis is the privileged axis within our convention for the Pauli matrix. The real operators $(a_k, \,\, b_k)$ are called Majorana operators. They may always be obtained independent on the system we are considering, because Eq.~(\ref{majopt}) is a generic transform. Hence, the Majorana number is essential to distinguish between a trivial or a  Majorana phases. For $M=+1$ the phase is trivial, but $M=-1$ there exist Majorana phase, which must be achieved  by tuning the parameters of the Hamiltonian. We shall apply this concept for the Kitaev Chain in Section A.~V. 

\subsection{The Two Dimensional Case}

For two-dimensional systems, we return to Eq.~(\ref{bermag}) in order to calculate the Berry phase. The first step is calculate the gradient of $H$ in the h-space. From Eq.~(\ref{h2}), we obtain
\begin{equation}
\nabla_h H=\vec{\sigma}.
\end{equation}

We remember that $\vec{\sigma}$ is the pseudo-spin of the system. It is the degree of freedom generated by the sublattices of the system, similar to the spin of the electron. Therefore, we may consider that the vector $\textbf{h}$ points along the $h_z$-axis, i.e, we consider $h_1=h_2=0$ in Eq.~(\ref{h2}). In this rotated axis, the eigenvectors are
\begin{equation}
|+ \rangle=\left(\begin{matrix} 1 \\ 0 \end{matrix}\right),|- \rangle=\left(\begin{matrix} 0 \\ 1 \end{matrix}\right),
\end{equation}
satisfying the relations
\begin{equation}
\sigma_z|\pm \rangle=\pm |\pm \rangle , \sigma_x |\pm \rangle= |\mp \rangle, \sigma_y |\pm \rangle=\pm i|\mp \rangle.
\end{equation}

Using Eq.~(\ref{bermag}), we may find the vector $\textbf{V}_l$ for $l=+$, which reads
\begin{eqnarray}
\textbf{V}_{+}&=&-\textbf{Im}\frac{\langle +| \vec{\sigma} |-\rangle \times \langle -|\vec{\sigma}|+\rangle}{4 |h|^2}\nonumber\\
&=&-\textbf{Im}\frac{(\hat h_x+i\hat h_y)\times (\hat h_x-i\hat h_y)}{4|h|^2}\nonumber\\
&=&-\frac{\hat h_z}{2|h|^2}.
\end{eqnarray}
For $l=-$, we obtain $\textbf{V}_-=-\textbf{V}_+$. Using translational invariance in the h-space, we have the coordinate-independent expression for this vector
\begin{equation}
\textbf{V}_+=-\frac{\textbf{h}}{2|h|^3}. \label{v+}
\end{equation}

The final step is to calculate the integral over the surface of the vector $\textbf{V}_+$. Next, we obtain the Berry phase $\gamma_l$ for $l=-$
\begin{equation}
\gamma_-=\int_S d\textbf{S}_h. \frac{\textbf{h}}{2|h|^3}=\frac{\Omega_h}{2}. \label{solang}
\end{equation}

Remarkably, from Eq.~(\ref{solang}) we have shown that, for two-dimensional systems, the Berry phase is half of the solid angle $\Omega_h$ swept out by the unit vector $\hat{\textbf{h}}(k)$. Nevertheless, Eq.~(\ref{solang}) may be rewritten as an integral in the momentum space. Indeed, first we note that surface $d\textbf{S}_h$ is
\begin{equation}
d\textbf{S}_h=dh_2dh_3 \hat{h}_1+dh_1dh_3 \hat{h}_2+dh_1dh_2 \hat{h}_3. \label{surele}
\end{equation}
Thereafter, we convert these surface integral from the h-space to the k-space, using the Jacobian transformation $J(h\rightarrow k)$. For instance,
\begin{equation}
dh_2dh_3=dk_xdk_y J(h_2,h_3\rightarrow k_x,k_y),
\end{equation}
where
\begin{equation}
J(h_2,h_3\rightarrow k_x,k_y)=\det \left(\begin{matrix} \frac{\partial h_2}{\partial k_x}& \frac{\partial h_2}{\partial k_y} \\ \frac{\partial h_3}{\partial k_x}&\frac{\partial h_3}{\partial k_y} \end{matrix}\right)=\left(\frac{\partial \textbf{h}}{\partial k_x}\times \frac{\partial \textbf{h}}{\partial k_y}\right)_{\hat{h}_1}
\end{equation}
is the corresponding Jacobian to this transformation. Thereafter, we perform this transformation for all terms in Eq.~(\ref{surele}) and use this result into Eq.~(\ref{solang}), we find
\begin{equation}
\gamma_-=\frac{1}{2}\int_{B.Z}\frac{\textbf{h}}{|h|^3}.\left(\frac{\partial \textbf{h}}{\partial k_x}\times \frac{\partial \textbf{h}}{\partial k_y}\right)dk_x dk_y.
\end{equation}

Similar to the one dimensional problem, we define a topological invariant $n$, the Cher number, which is the Berry phase divided by $2\pi$. The practical expression for the Chern number is \cite{Bernevig}
\begin{equation}
n=\frac{1}{4\pi}\int_{B.Z}\hat{\textbf{h}}.\left(\frac{\partial \hat{\textbf{h}}}{\partial k_x}\times \frac{\partial \hat{\textbf{h}}}{\partial k_y}\right)dk_x  dk_y.  \label{chn}
\end{equation}

The only hurdle to calculate the Chern number is to perform the k-integral in Eq.~(\ref{chn}). Remarkably, this is not necessary as it has been pointed out by authors in Ref.~\cite{geochn}. Here, we just summarize their results in order to expose a very practical procedure to calculate the Chern number. They have converted the integral in Eq.~(\ref{chn}) into a sum over high-symmetry points or also called Dirac points, therefore,
\begin{equation}
n=\frac{1}{2}\sum_{\textbf{k} \in D_i} \sgn (J_i) \sgn (h_i), \label{geochn}
\end{equation}
where $h_i$ is some privileged axis and $D_i$ are the high-symmetry points. $J_i$ is the Jacobian calculated in the $D_i$ points, it is given by
\begin{equation}
J_i=(\partial_{k_x} \hat{h} \times \partial_{k_y} \hat{h})_i. \label{jac}
\end{equation}

There exist some conditions to apply Eq.~(\ref{geochn}) for calculating the Chern number. Indeed, the $D_i$ points are obtained by setting two axis to zero, for instance $h_1(D_i)=h_2(D_i)=0$, therefore $h_3$ is the privileged axis with $h_3(D_i)\neq 0$. The Jacobian $J_i$ must be nontrivial when calculated in $D_i$. In this scheme, the chirality $J_3$ is defined as the sign of the Jacobian \cite{geochn}, therefore,
\begin{equation}
J_3=\sgn(J_i).
\end{equation}

We shall calculate the winding and Chern numbers for some modes. Nevertheless, it is insightful to understand what means a nontrivial Berry phase for a two-dimensional system. In order to do so, we derive the quantum adiabatic theorem \cite{Bernevig} in Section A.~III. Thereafter, we apply this result in Section A.~VI to show its relation with the Hall conductivity. 

\section{A.III-The quantum adiabatic theorem.}
\label{cap53} 

We use the full series of the coefficients $a_n(t)$ in Eq.~(\ref{exppsi02}) into Eq.~(\ref{eqsch00x}). Thereby, we find
\begin{equation}
\sum_n e^{-i\theta_n(t)/\hbar} \frac{\partial a_n(t)}{\partial t} |n(t) \rangle=-\sum_m a_m(t) \frac{\partial |m(t) \rangle}{\partial t} e^{-i\theta_m(t)/\hbar}.
\end{equation}

Using $\langle l(t)|n(t) \rangle=\delta_{lm}$, we have
\begin{equation}
\dot{a}_l(t)=-\sum_m a_m(t) \langle l(t)|\frac{\partial}{\partial t}|m(t) \rangle \, e^{-i(\theta_m(t)-\theta_l(t))/\hbar}. \label{an0}
\end{equation}

There are two terms for $m=l$ and $m \neq l$. Therefore,
\begin{eqnarray}
\dot{a}_l(t)&=&-a_l(t)\langle l(t)|\frac{\partial}{\partial t}|l(t)\rangle \nonumber\\
&-&\sum_{m\neq l} a_m(t) \langle l(t)|\frac{\partial}{\partial t}|m(t) \rangle \, e^{-i(\theta_m(t)-\theta_l(t))/\hbar}. \label{eqrecori}
\end{eqnarray}

This is the best we can do without any approximations. Note that $\langle l(t)|\frac{\partial}{\partial t}|l(t)\rangle=\dot{\textbf{k}}.\langle l(\textbf{k})|\nabla_{\textbf{k}}|l(\textbf{k})\rangle$. Hence, According to Eq.~(\ref{an0}), for $\dot{\textbf{k}}= 0$, we have $\dot{a}_l(t)= 0$, i.e, $a_l(t)=a_l(0)$. This is the quantum adiabatic theorem in its lowest order that we have applied in Section A.~I. We shall relax the previous condition by keeping the linear power of $\dot{\textbf{k}}$ in $a_n(t)$. We separate the lowest-order solution from the corrections, thereby,  $a_l(t)=a_l(0)+\sum_{i=1}^{\infty} a^{(i)}_l(t)$, where $ a^{(i)}_l(t) \propto O((\dot{\textbf{k}})^i)$. Using this decomposition in Eq.~(\ref{eqrecori}), we have  
\begin{eqnarray}
\dot{a}_l(t)&=&-\sum_{m\neq l} (a_m(0)+\sum_{i=1}^{\infty} a^{(i)}_m(t)) \langle l(t)|\frac{\partial}{\partial t}|m(t) \rangle \nonumber\\ \, && e^{-i(\theta_m(t)-\theta_l(t))/\hbar}, \label{eqrecor2}
\end{eqnarray}
where we have neglected the first term in Eq.~(\ref{eqrecori}), because it does not represent a transition between different eigenstates.

Using $a_l(t)=a_l(0)+a^{(1)}_l(t)$ in Eq.~(\ref{eqrecor2}), we find
\begin{eqnarray}
\dot{a}^{(1)}_l(t)&=&-\sum_{m\neq l} a_m(0) \langle l(t)|\frac{\partial}{\partial t}|m(t) \rangle \nonumber \\
&&\, e^{-i(\theta_m(t)-\theta_l(t))/\hbar}. \label{eqrecor25}
\end{eqnarray}
Integrating out the time, we have
\begin{eqnarray}
a^{(1)}_l(t)-a^{(1)}_l(0) &=& -\sum_{m\neq l}\int_{0}^t dt' a_m(0) \langle l(t')|\frac{\partial}{\partial t'}|m(t') \rangle \nonumber\\ && e^{-i(\theta_m(t')-\theta_l(t'))/\hbar}. \label{eqrecor3}
\end{eqnarray}
Since $\langle n(t')|\frac{\partial}{\partial t'}|m(t') \rangle \propto \dot{\textbf{k}}<<1$, therefore, integration over this term yields $\dot{\textbf{k}}^2$-proportional terms, which we are neglecting from the very beginning. For the sake of consistency, we integrate only over the frequencies $\theta_n(t)$. Therefore, using Eq.~(\ref{propri}) in Eq.~(\ref{eqrecor3}), we find  
\begin{eqnarray}
a^{(1)}_l(\textbf{k})&=& a^{(1)}_l(0)-i\hbar\,\sum_{m\neq l}a_m(0)\, \dot{\textbf{k}}. \frac{\langle l(\textbf{k})| (\nabla_{\textbf{k}}H)|m(\textbf{k}) \rangle}{E_m(\textbf{k})-E_l(\textbf{k})} \nonumber\\
&& e^{-i(\theta_m(\textbf{k})-\theta_l(\textbf{k}))/\hbar}.  \label{afinal}
\end{eqnarray}

The phase factors in Eq.~(\ref{afinal}) may be neglected with a simple redefinition of the $l$ and $m$-states. Therefore, using Eq.~(\ref{afinal}) in Eq.~(\ref{exppsi02}), we have
\begin{equation}
|\Psi(\textbf{k}) \rangle=\sum_n (a_n(0)+a^{(1)}_n(\textbf{k})+O(\dot{\textbf{k}}^2))|n(t) \rangle,
\end{equation}
where the sum over $n$ reads
\begin{eqnarray}
|\Psi(\textbf{k}) \rangle&=& a_l(0)|l(\textbf{k}) \rangle+ \sum_{n\neq l} a_n(0)|n(\textbf{k}) \rangle +a^{(1)}_l(\textbf{k})|l(\textbf{k}) \rangle\nonumber\\&+&\sum_{n\neq l} a^{(1)}_n(\textbf{k}) |n(\textbf{k}) \rangle.
\end{eqnarray}

Let us choose $a_n(0)=\delta_{nl}$, i.e, we perform a projection from $|\Psi(\textbf{k}) \rangle$ to $|\psi_l(\textbf{k}) \rangle$. From Eq.~(\ref{afinal}), we have $a^{(1)}_l(\textbf{k})=a^{(1)}_l(0)$. For simplicity, $a^{(1)}_l(0)=0$, then
\begin{equation}
|\psi_l(\textbf{k}) \rangle= |l(\textbf{k}) \rangle +\sum_{n\neq l} a^{(1)}_n(\textbf{k}) |n(\textbf{k}) \rangle. \label{inter}
\end{equation}

Using Eq.~(\ref{afinal}) in Eq.~(\ref{inter}), we have
\begin{eqnarray}
|\psi_l(\textbf{k}) \rangle &=& |l(\textbf{k}) \rangle -i\hbar \,\sum_{n\neq l}\,\sum_{m\neq n}a_m(0)\, \dot{\textbf{k}} \nonumber \\
&.&\frac{\langle n(\textbf{k})| (\nabla_{\textbf{k}}H)|m(\textbf{k}) \rangle}{E_m(\textbf{k})-E_n(\textbf{k})} |n(\textbf{k}) \rangle,
\end{eqnarray}
where we have imposed $a_m(0)=\delta_{ml}$. Finally, we find 
\begin{equation}
|\psi_l(\textbf{k}) \rangle = |l(\textbf{k}) \rangle -i\hbar \sum_{n\neq l}  |n(\textbf{k}) \rangle\, \dot{\textbf{k}}.  \frac{\langle n(\textbf{k})| (\nabla_{\textbf{k}}H)|l(\textbf{k}) \rangle}{E_l(\textbf{k})-E_n(\textbf{k})}. \label{taq}
\end{equation}

Eq.~(\ref{taq}) is the quantum adiabatic theorem \cite{Bernevig}. We shall use this result in the next section to show the relation between the Berry phase and the Hall conductivity.

\section{A.~IV-The Berry Phase and the Hall conductivity.}
\label{cap55} 

In this section, we show the relation between the Berry phase and the Hall conductivity. In order to do so, we shall use the quantum adiabatic theorem we have obtained in Eq.~(\ref{taq}). A particle with charge $-e$ and momentum $\hbar \textbf{k}$ in the presence of an external electric field $\textbf{E}^{(e)}$ satisfies the dynamical equation 
\begin{equation}
\hbar \dot{\textbf{k}}=-e \textbf{E}^{(e)}.
\end{equation}

The average velocity in the $l$-state is given by
\begin{equation}
_l\langle \textbf{v}(\textbf{k}) \rangle_l=\langle \psi_l(\textbf{k}) | \textbf{v}(\textbf{k})|\psi_l(\textbf{k}) \rangle =\langle \psi_l(\textbf{k}) | \frac{\partial H}{\hbar \partial \textbf{k}}|\psi_l(\textbf{k}) \rangle. \label{avvel}
\end{equation}
Using Eq.~(\ref{taq}) in Eq.~(\ref{avvel}), we find
\begin{eqnarray}
_l\langle \textbf{v}(\textbf{k}) \rangle_l&=& \langle \psi_l(\textbf{k}) | \frac{\partial H}{ \hbar \partial \textbf{k}}|\psi_l(\textbf{k})\rangle-i\frac{e}{\hbar}\sum_{n\neq l} E^{(e)}_i \nonumber\\
&&\left[\frac{\langle l(\textbf{k})|\frac{\partial H}{ \partial k_i} |n(\textbf{k})\rangle \langle n(\textbf{k})|\frac{\partial H}{ \partial \textbf{k}} |l(\textbf{k})\rangle }{(E_l(\textbf{k})-E_n(\textbf{k}))^2}-h.c\right]. \label{avvel0}
\end{eqnarray}

For consistency, we have neglected the terms in order of $E{^{(e)}}^2$. Eq.~(\ref{avvel0}) may be written in a insightful expression given by     
\begin{eqnarray}
_l\langle \textbf{v}(\textbf{k}) \rangle_l&=&\langle \psi_l(\textbf{k}) | \frac{\partial H}{\hbar  \partial \textbf{k}}|\psi_l(\textbf{k})\rangle-i\frac{e}{\hbar}\sum_{n\neq l} \nonumber
\\
&&\left( \frac{\langle n(\textbf{k})|\frac{\partial H}{ \partial k_y} |l(\textbf{k})\rangle \langle l(\textbf{k})|\frac{\partial H}{ \partial k_x} |n(\textbf{k})\rangle}{(E_l(\textbf{k})-E_n(\textbf{k}))^2}-h.c\right)\nonumber\\&&
\left(\begin{matrix} -E^{(e)}_y \\ E^{(e)}_x \end{matrix}\right). \label{avvel1}
\end{eqnarray}

The electric current in the system is
\begin{equation}
\textbf{j}=-e \sum_l\, \int \frac{d^2 \textbf{k}}{(2\pi)^2}\, _l\langle \textbf{v}(\textbf{k}) \rangle_l f(E_{l,k}), \label{jcurele}
\end{equation}
where $f(E_{l,k})=1/(e^{\beta (E_{l,k}-\mu)}+1)$ is the Fermi-Dirac function, $\beta=1/(k_B T)$, $k_B$ is the Boltzmann's constant, $T$ is temperature, $\mu(T)$ is the chemical potential, and $E_{l,k}$ is the energy of the $l$-state, which is dependent on the momentum $k$. For $T\rightarrow 0$, we have $f(E_{l,k})=\Theta(E_{l,k}-E_F)$, where $\Theta(x)=1$ when $x>0$, $\Theta(x)=0$ for $x<0$, is the step function. Furthermore, $E_F=\mu(T=0)$ is the Fermi energy.
This limit implies that the sum over $l$ and the integral over $\textbf{k}$ must be restricted only to the set $l_0$ of filled bands.
Therefore, Eq.~(\ref{jcurele}) becomes
\begin{equation}
\textbf{j}=\left(\begin{matrix} \sigma_{xx}& \sigma_{xy} \\ \sigma_{yx}&\sigma_{yy} \end{matrix}\right) \left(\begin{matrix} E^{(e)}_x \\ E^{(e)}_y \end{matrix} \right),
\end{equation}
using Eq.~(\ref{avvel1}), we obtain the conductivities $\sigma_{xx}=\sigma_{yy}$, $\sigma_{xy}=-\sigma_{yx}$, where
\begin{equation}
\sigma_{xx}=-e\,\sum_{l \in l_0}\, \int \frac{d^2 \textbf{k}}{(2\pi)^2} \langle \psi_l(\textbf{k}) | \frac{\partial H}{\hbar  \partial k_x}|\psi_l(\textbf{k}) \rangle
\end{equation}
is called the longitudinal conductivity and
\begin{eqnarray}
\sigma_{xy}&=&-i\frac{e^2}{h}\frac{1}{2\pi}\,\sum_{l \in l_0}\, \int d^2 \textbf{k}\sum_{n\neq l} \nonumber \\
&&
\left( \frac{\langle n(\textbf{k})|\frac{\partial H}{ \partial k_y} |l(\textbf{k})\rangle \langle l(\textbf{k})|\frac{\partial H}{ \partial k_x} |n(\textbf{k})\rangle}{(E_l(\textbf{k})-E_n(\textbf{k}))^2}-h.c\right) \label{curaux}
\end{eqnarray}
is called the Hall conductivity. 

Using Eq.~(\ref{propri}), we have
\begin{equation}
\frac{\langle n(\textbf{k})|\frac{\partial H}{\partial k_y} |l(\textbf{k}) \rangle}{E_l-E_n}=\langle n(\textbf{k}) |\partial_{k_y}l(\textbf{k}) \rangle
\end{equation}
and
\begin{equation}
\frac{\langle l(\textbf{k})|\frac{\partial H}{\partial k_x} |n(\textbf{k}) \rangle}{E_l-E_n}=-\langle l(\textbf{k}) |\partial_{k_x}n(\textbf{k}) \rangle=\langle \partial_{k_x} l(\textbf{k}) |n(\textbf{k}) \rangle.
\end{equation}
Therefore, Eq.~(\ref{curaux}) becomes
\begin{eqnarray}
\sigma_{xy}&=&-i\frac{e^2}{h}\frac{1}{2\pi}\,\sum_{l \in l_0}\, \int d^2 \textbf{k}\nonumber \\
&&\sum_{n\neq l} \langle \partial_{k_x} l(\textbf{k}) |n(\textbf{k}) \rangle \langle n(\textbf{k}) |\partial_{k_y}l(\textbf{k})\rangle -h.c.
\end{eqnarray}
We have to keep only the real part of $\sigma_{xy}$. Therefore,
\begin{eqnarray}
\sigma_{xy}&=&-\textbf{Im} \frac{e^2}{h}\frac{1}{2\pi}\,\sum_{l \in l_0}\, \int d^2 \textbf{k}\sum_{n\neq l} \nonumber \\ &&
\langle \partial_{k_x} l(\textbf{k}) |n(\textbf{k}) \rangle \langle n(\textbf{k}) |\partial_{k_y}l(\textbf{k})\rangle -h.c.
\end{eqnarray}
We may include the term $\langle \partial_{k_x} l(\textbf{k}) |l(\textbf{k}) \rangle \langle l(\textbf{k}) |\partial_{k_y}l(\textbf{k})\rangle -h.c$, thereby
\begin{eqnarray}
\sigma_{xy}&=&-\textbf{Im}  \frac{e^2}{h}\frac{1}{2\pi}\,\sum_{l \in l_0}\, \int d^2 \textbf{k}\sum_{n} \nonumber \\ && \langle \partial_{k_x} l(\textbf{k}) |n(\textbf{k}) \rangle \langle n(\textbf{k}) |\partial_{k_y}l(\textbf{k})\rangle -h.c.
\end{eqnarray}

Next, we solve the sum over $n$, yielding 
\begin{equation}
\sigma_{xy}=-\frac{e^2}{h}\frac{1}{2\pi}\times \sum_{l \in l_0}\textbf{Im}  \, \int dk_xdk_y \langle \partial_{k_x} l(\textbf{k}) |\partial_{k_y}l(\textbf{k})\rangle -h.c . \label{hallber0}
\end{equation}

Remarkably, from comparison with Eq.~(\ref{berphcur}) and Eq.~(\ref{hallber0}), we conclude that
\begin{equation}
\sigma_{xy}=\frac{e^2}{h} \frac{1}{2\pi} \sum_{l \in l_0} \gamma^l, \label{chnhall}
\end{equation}
where
\begin{equation}
\gamma^l=-\textbf{Im}\left\{\int_{B.Z} dk_x dk_y\, \Omega^l_{k_xk_y}\right\}. \label{chnhall2}
\end{equation}

This result is the most important consequence of the nontrivial topology in a two-dimensional system. It shows that when the sum of the Berry phase of all filled bands is not zero, then the material has a nonzero Hall conductivity, whose quantized coefficient is the Chern number. This result has been shown in terms of the Kubo formula in Ref.~\cite{zhangtop}.

\section{A.~V-Topological phases in the SSH and Kitaev models.}
\label{cap54} 

In this section, we describe the topological phases for two important models in one dimension, namely, the Su-Schrieffer-Hegger (SSH) and Kitaev models. We start with the SSH model.

\subsection{SSH Model}

\begin{figure}[htb]
\label{SSHf}
\centering
\includegraphics[scale=0.5]{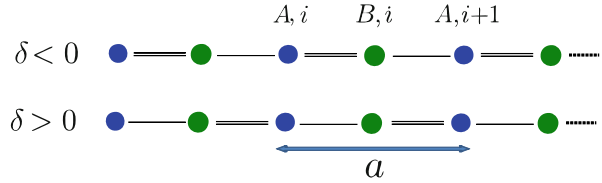}
\caption{The SSH model in real space for $t>0$. $a\equiv 1$ is the lattice parameter. The two phases are connected by the discrete transformation $\delta\rightarrow -\delta$. The double and single lines represent the long and short hopping parameter, respectively.} \label{figssh1}
\end{figure}

The Hamiltonian of the SSH model is
\begin{eqnarray}
H&=&-\sum_{n=1}^{N}(t+\delta)[(c^\dagger_{An}c_{Bn}+h.c)\nonumber \\ &+&(t-\delta)(c^\dagger_{An+1}c_{Bn}+h.c)], \label{SSH}
\end{eqnarray}
where $n_{in}=c^\dagger_{in}c_{in}$ is the particle number operator, $t+\delta$ is called the long hopping parameter, $t-\delta$ is the short hopping parameter, and $(A, B)$ are the two sublattices. The transformation $\delta\rightarrow -\delta$ changes the phase of the system, because the long hopping parameters become the short one. In Fig.~\ref{figssh1}, we illustrate the two phases one with $\delta<0$ and the other with $\delta>0$. A priori, these phases are equivalent. Nevertheless, they have different topology for open boundary conditions, because they have different topological invariants. We shall prove this result, using the Berry phase.

We would like to calculate the Berry phase, using Eq.~(\ref{ber1d}). In order to do so, we have to obtain the Bloch Hamiltonian that gives the function $h(k)$. This is obtained by the Fourier transform. The Fourier transform of the annihilation operators for the $A$ and $B$ are
\begin{equation}
a_k=\frac{1}{\sqrt{N}}\sum_n \exp(-ikna)c_{An} \label{akSSH}
\end{equation}
and
\begin{equation}
b_k=\frac{1}{\sqrt{N}}\sum_n \exp(-ikna)c_{Bn}. \label{bkSSH}
\end{equation}
Furthermore, the orthogonality of exponential function reads
\begin{equation}
\sum_n e^{-ina(k-q)}=N\delta_{kq}. \label{ortprop}
\end{equation}

Using Eq.~(\ref{akSSH}), Eq.~(\ref{bkSSH}), and Eq.~(\ref{ortprop}) into Eq.~(\ref{SSH}), we find the Bloch Hamiltonian of the SSH model, given by
\begin{equation}
H=\sum_k \psi^\dagger_k(d_x \sigma_x+d_y \sigma_y)\psi_k, \label{BSSH}
\end{equation}
where $\psi^\dagger_k=(a_k\,\,\,b_k)$ is a spinor field, $\sigma_x$ and $\sigma_y$ are the usual Pauli matrix,
\begin{equation}
d_x=(t+\delta)+(t-\delta) \cos k, \label{dxSSH}
\end{equation}
and
\begin{equation}
d_y=(t-\delta) \sin k. \label{dySSH}
\end{equation}

For the sake of simplicity, we have considered $a=1$. Remarkably, in the low-energy limit $k\rightarrow 0$, we have $(d_x,d_y)=(2t,(t-\delta)k)$, which is, essentially, the Dirac theory in (1+1) dimensions.

The eigenvalues of the Bloch Hamiltonian in Eq.~(\ref{BSSH}) are
\begin{equation}
E_{\pm}(k)=\pm\sqrt{d_x^2+d_y^2}=\pm2\sqrt{t^2\cos^2\left(\frac{k}{2}\right)+\delta^2\sin^2\left(\frac{k}{2}\right)}. \label{ESSH}
\end{equation}
\begin{figure}[htb]
\label{SSHenergy}
\centering
\includegraphics[scale=0.8]{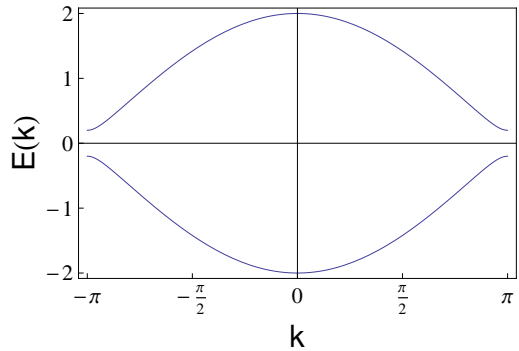}
\caption{The energy dispersion for the SSH model calculated in Eq.~(\ref{ESSH}) for $\delta=0.1$, $t=1.0$, and $k\in[-\pi,+\pi]$. The energy gap at the points $k=\pm \pi$ is $2|\delta|$.} \label{fig2}
\end{figure}
In Fig.~\ref{fig2}, we plot Eq.~(\ref{ESSH}). At the point $k=\pi$, the energy gap is $2\delta$. Therefore, the gap only closes if $\delta\rightarrow 0$. In this case, we break the adiabatic condition in the Hamiltonian, hence there exist a topological phase transition.

For the SSH model, the winding number is given by Eq.~(\ref{ber1d}). Thereby,
\begin{equation}
\nu_{{\rm SSH}}=-\frac{i}{2\pi}\int_{B.Z} dk\,\hat d^{-1}(k)\partial_k \hat d(k), \label{nuSSH}
\end{equation}
where $\hat d(k)\equiv d(k)/|d(k)|$ with $d(k)=d_x(k)+id_y(k)$, given by Eq.~(\ref{BSSH}). In general grounds, because the integral in Eq.~(\ref{nuSSH}) is over the first Brillouin zone $k \in [-\pi,+\pi]$, the winding reads
\begin{equation}
\nu_{{\rm SSH}}=-\frac{i}{2\pi}\int_{B.Z} dk\, d^{-1}(k)\partial_k \ d(k). \label{nuSSH2}
\end{equation}

After some simplifications, we find
\begin{equation}
\nu_{{\rm SSH}}=-\frac{i}{2\pi}\int_{B.Z} dk {\cal K}(k, t,\delta) \label{nuSSH3}
\end{equation}
with
\begin{equation}
{\cal K}=i\frac{(t-\delta)^2 +(t^2-\delta^2)\cos k}{[t+\delta+(t-\delta)\cos k]^2+(t-\delta)^2\sin^2 k}.
\end{equation}

Integrating out $k$ in Eq.~(\ref{nuSSH3}) for $t>0$, we have
\begin{equation}
\nu_{{\rm SSH}}=\frac{1}{2}\left(1-\sgn \delta\right).
\end{equation}

Therefore, the winding number of the SSH model is
\begin{equation}
\nu_{{\rm SSH}}=\left\{\begin{array}{rc}
1, & \delta<0,\\
0, & \delta>0. \\ 
\end{array}\right. \label{chnSSH}
\end{equation}
\begin{figure}[htb]
\label{sshbb}
\centering
\includegraphics[scale=0.45]{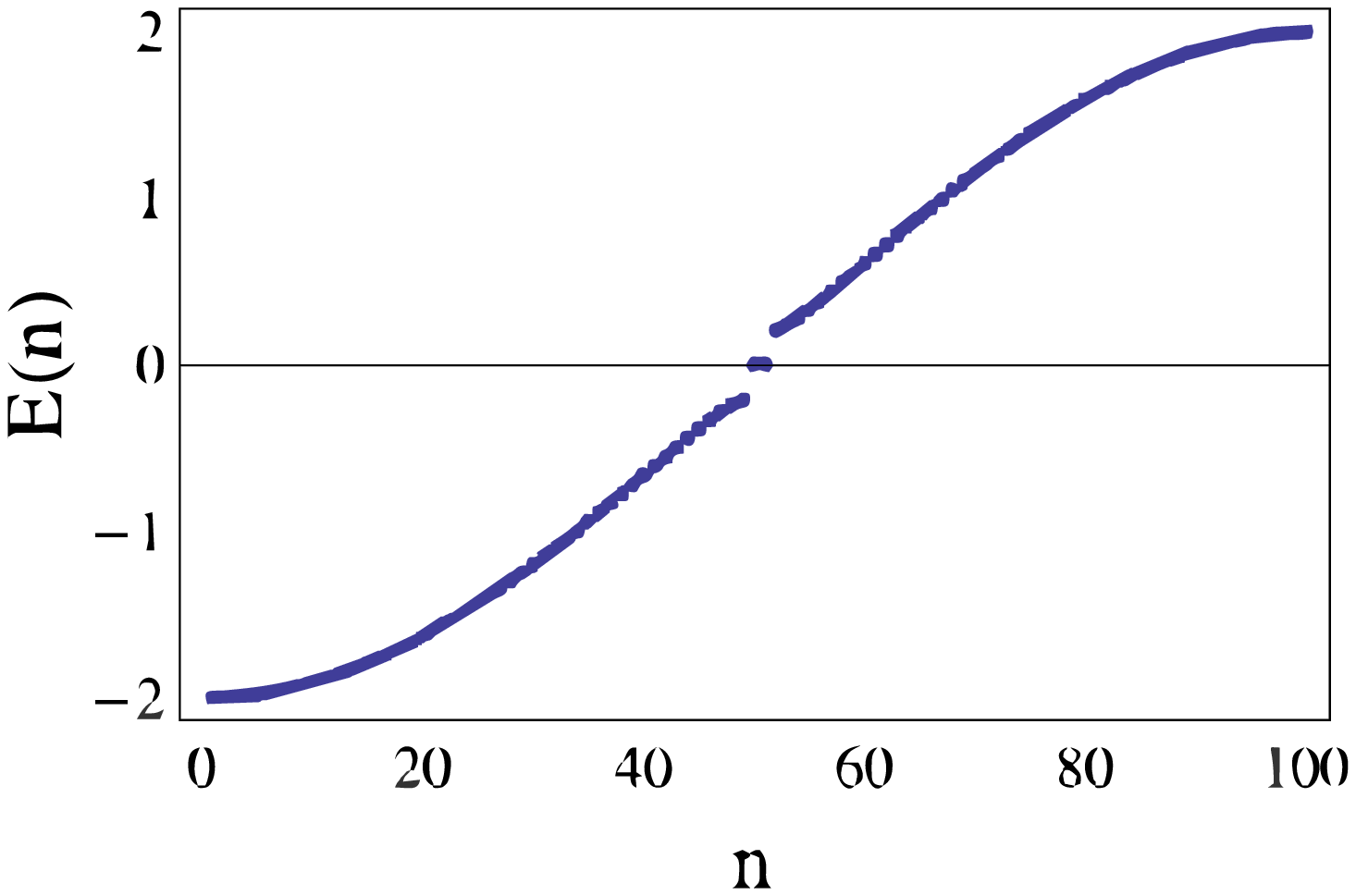}
\caption{The eigenvalues of the SSH model for $N=100$, $t=1.0$, and $\delta=-0.1$. } \label{fig3}
\end{figure}

\begin{figure}[htb]
\label{sshbb}
\centering
\includegraphics[scale=0.45]{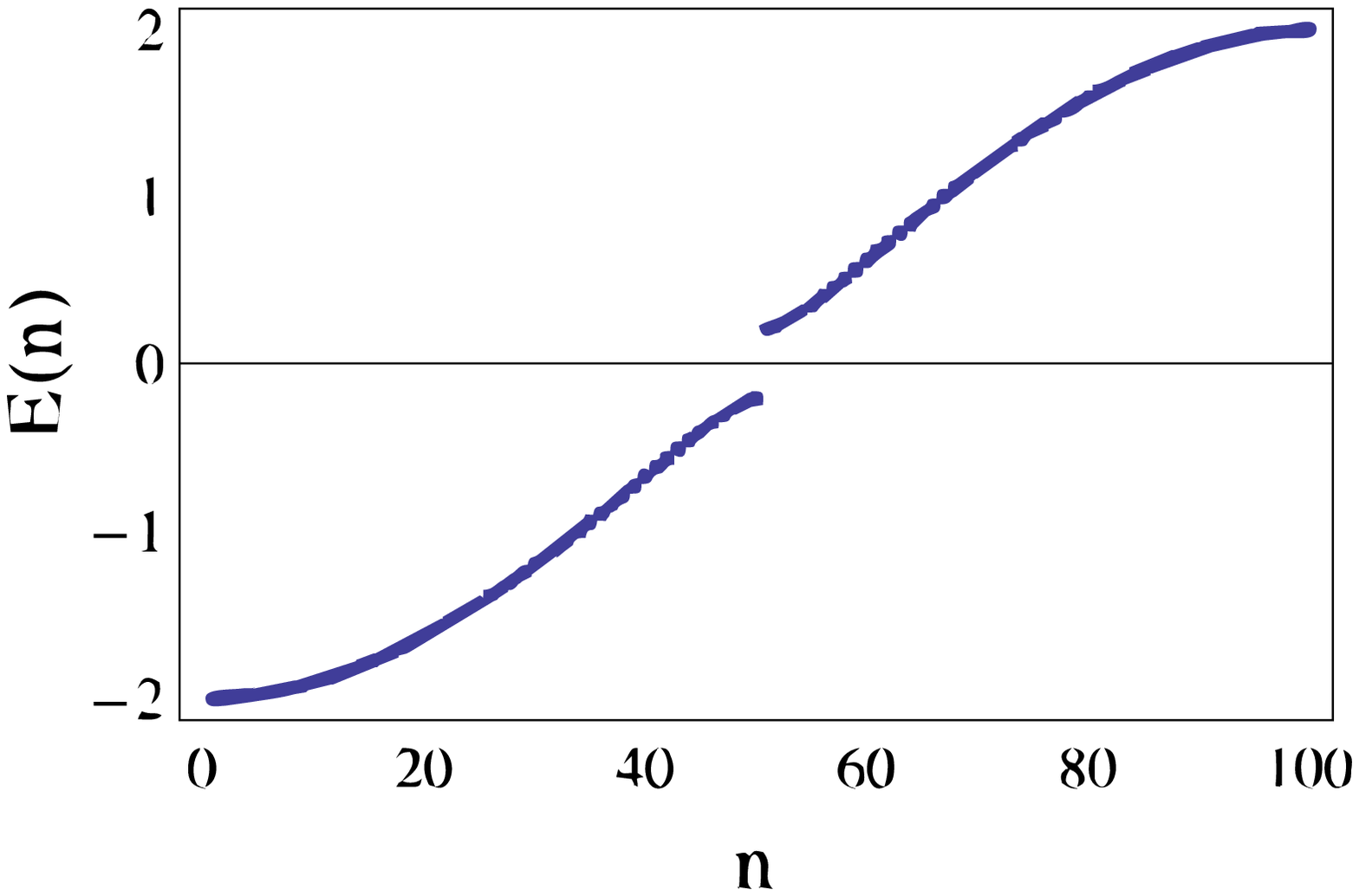}
\caption{The eigenvalues of the SSH model for $N=100$, $t=1.0$, and $\delta=+0.1$ } \label{fig4}
\end{figure}

The phase with nonzero winding number has zero-energy states. Indeed, it is easy to show that for $N=4$, the Hamiltonian in Eq.~(\ref{SSH}) reads
\begin{equation}
H={\cal C}^\dagger \left(\begin{matrix} 0 & -(t+\delta) & 0& 0 \\ -(t+\delta) &0 & -(t-\delta) &0 \\0& -(t-\delta)& 0& -(t+\delta) \\ 0&0& -(t+\delta)& 0  \end{matrix}\right) {\cal C}, \label{sshsite}
\end{equation}
where ${\cal C}^\dagger=(c^\dagger_{A,1} \,\, c^\dagger_{B,1} \,\, c^\dagger_{A,2} \,\, c^\dagger_{B,2})$ is the basis. The generalization for any $N$ is straightforward. We have to calculate the eigenvalues of Eq.~({\ref{sshsite}}). In Fig.~\ref{fig3} and Eq.~\ref{fig4}, we plot the eigenvalues for each site $n$ with $N=100$, and $t=1.0$. In the Fig.~\ref{fig3}, we assume $\delta=-0.1$, thereby we have obtained that the SSH model has zero-energy states. In Fig.~\ref{fig4}, we assume $\delta=0.1$, then we obtained that the system has no zero-energy states.

The existence of zero-energy modes implies the existence of a ground state with fractional charge \cite{Jackiw}. Indeed, we write the decomposition of the field operator as
\begin{equation}
\psi(x)=\sum_E b_E \psi_E(x)+d^\dagger_E\psi_{-E}(x)+a\psi_0(x), \label{fieldoperator}
\end{equation}
where $b_E$ creates an electron with positive energy, $d^\dagger_E$ creates a hole with negative energy, and $a$ creates the zero-energy mode. These fermionic operators obey: $\{b_E,b_E^\dagger\}=1$, $\{d_E,d_E^\dagger\}=1$, and $\{a,a^\dagger\}=1$. 

From the particle-hole symmetry, we conclude that for each sate with energy $E$, there is other state with energy $-E$. Therefore, the zero-energy mode $E=0$ must be double degenerate. We represent these zero-energy states by $|G_\pm \rangle$, hence, $\{a,a^\dagger\}|G_\pm \rangle=|G_\pm \rangle$. This allow us to obtain the algebra of these operators given by
\begin{equation}
a|G_+ \rangle=|G_- \rangle,\,a|G_- \rangle=0,\,a^\dagger|G_- \rangle=|G_+ \rangle,\,a^\dagger|G_+ \rangle=0. \label{algebra}
\end{equation}

Next, we define the fermion number operator $Q$. This is given by
\begin{equation}
Q=\frac{1}{2}\int d^2 x(\psi^\dagger \psi-\psi \psi^\dagger). \label{charge}
\end{equation}
Using Eq.~(\ref{fieldoperator}) in Eq.~(\ref{charge}), we find
\begin{equation}
Q=\sum_E(b^\dagger_E b_E-d^\dagger_E d_E)+a^\dagger a-\frac{1}{2}. \label{charge2}
\end{equation}

Using Eq.~(\ref{charge2}) and Eq.~(\ref{algebra}), it is easy to show that
\begin{equation}
Q|G_\pm \rangle=\pm \frac{1}{2}|G_\pm \rangle.
\end{equation}

The states with fractional quantum numbers are known as solitons.

\subsection{Kitaev Chain}

Here, we discuss the so-called Kitaev chain. Essentially, the model consists of a quantum wire placed on the surface of a 3-dimensional superconductor. The Hamiltonian of the Kitaev chain reads
\begin{eqnarray}
H&=&-\sum_{j=1}^{N-1}\left[t(c^\dagger_jc_{j+1}+c^\dagger_{j+1}c_{j})-\frac{\Delta}{2}(c_jc_{j+1}+c^\dagger_{j+1}c^\dagger_j)\right] \nonumber\\&-&\mu\sum_{j=1}^N n_j,  \label{Kitaevchain}
\end{eqnarray} 
where the $\Delta$-term is due to the superconductor that breaks the Gauge symmetry $c_j\rightarrow c_j e^{i \phi}$, but preserves the discrete symmetry $c_j\rightarrow -c_j$. The $t$ and $\mu$ terms are the usual hopping parameter and chemical potential, respectively. $n_j=c^\dagger_j c_j$ is the particle number operator at the site $j$.

We apply the Fourier transform in the creation and annihilation operators, given by
\begin{equation}
c^\dagger_j=\frac{1}{\sqrt{N}}\sum_k \exp(-ik x_j) c^\dagger_k,\, k\in \left[-\frac{\pi}{a},+\frac{\pi}{a}\right], \, x_j=j a. \label{ckitaev}
\end{equation}
Using Eq.~(\ref{ckitaev}) in Eq.~(\ref{Kitaevchain}), we find
\begin{equation}
H=\sum_k \epsilon_k c^\dagger_k c_k+\frac{\Delta}{2} \sum_k \left(e^{ika} c_{-k}c_k+e^{-ika}c^\dagger_k c^\dagger_{-k}\right), \label{Kitaev2}
\end{equation}
where
\begin{equation}
\epsilon_k=-2t \cos(ka)-\mu.
\end{equation}
Since the sum over the $k$ is symmetric and $\epsilon_k=\epsilon_{-k}$, we have the identity
\begin{eqnarray}
\sum_k\epsilon_k c^\dagger_k c_k&=&\frac{1}{2}\sum_k\epsilon_k(c^\dagger_k c_k+c^\dagger_{-k}c_{-k})\nonumber \\&=&\frac{1}{2}\sum_k \epsilon_k(c^\dagger_k c_k-c_{-k}c^\dagger_{-k}+1). \label{Kprop}
\end{eqnarray}

Using Eq.~(\ref{Kprop}) into Eq.~(\ref{Kitaev2}), but neglecting the term $1/2\sum_k \epsilon_k$, which is just a overall constant, we obtain
\begin{equation}
H=\frac{1}{2}\sum_k {\cal C}^\dagger_k H_{kk} {\cal C}_k, \, {\cal C}_k=(c^\dagger_k \,\, c_{-k}),
\end{equation}
where $H_{kk}$ is the Bloch Hamiltonian of the Kitaev model, given by
\begin{equation}
H_{kk}=\vec{\sigma}. \textbf{d}(\textbf{k}), \label{Kibloch}
\end{equation}
where $\vec{\sigma}$ are the usual Pauli matrix and
\begin{equation}
\textbf{d}(\textbf{k})=(\Delta \cos(ka), \Delta \sin(ka), \epsilon_k).
\end{equation}

The eigenvalues of the Bloch Hamiltonian in Eq.~(\ref{Kibloch}) are
\begin{equation}
E(k,\Delta)=\pm\sqrt{\Delta^2+(2t \cos(ka)+\mu)^2}.  
\end{equation}
Without the superconductor, we have $E(\pm\pi/a,0)=\pm|2t+\mu|=\pm |2t-|\mu||$ with $\mu<0$. Therefore, if the system cross the point $2t=|\mu|$, the gap closes and topology changes. Nevertheless, contrary to the SSH model, the nontrivial phase of the Kitaev model implies in the existence  of Majorana fermions at the end of the chain. The Majorana operators $\eta_j$ and $\gamma_j$ at each site $j$ are
\begin{equation}
c_j=\frac{1}{2}\left(\eta_j+i\gamma_j\right), \label{Majodef}
\end{equation} 
where $(\eta_j,\gamma_j)$ are reals. Due to the anti-commutative properties $\{c^\dagger_j,c_k\}=\delta_{jk}$ and $\{c_j,c_k\}=0$, we have
\begin{equation}
\{\gamma_j,\eta_k\}=0, \, \{\eta_j,\eta_k\}=2 \delta_{jk}, \, \{\gamma_j,\gamma_k\}=2\delta_{jk}. \label{Majoprop}
\end{equation}
The main goal is to convert the Kitaev Hamiltonian in Eq.~(\ref{Kitaevchain}), using Eq.~(\ref{Majodef}) and Eq.~(\ref{Majoprop}), from the electron operators to Majorana operators. This procedure may be done for any fermionic model.

The $t$-proportional term in Eq.~(\ref{Kitaevchain}), i.e, $c^\dagger_jc_{j+1}+h.c$  reads
\begin{eqnarray}
&& \frac{1}{4}(\eta_j+i \gamma_j)(\eta_{j+1}-i\gamma_{j+1})+h.c \nonumber= \\
&&\frac{1}{4}\left[i\gamma_j \eta_{j+1}-i\eta_{j+1}\gamma_j-i\eta_j\gamma_{j+1}+i\gamma_{j+1}\eta_{j}\right]= \nonumber \\
&&\frac{1}{4}\left[i \{\gamma_j,\eta_{j+1}\}-2i\eta_{j+1}\gamma_j-i\{\eta_j,\gamma_{j+1}\}+2i\gamma_{j+1}\eta_j\right]= \nonumber \\
&&\frac{i}{2}\left[\gamma_{j+1}\eta_j-\eta_{j+1}\gamma_j\right]. \label{ttermKi}
\end{eqnarray}

We have used in the rhs of Eq.~(\ref{ttermKi}) the property $\{A,B\}=[A,B]+2B A$, which is true for any $A$ and $B$ operators. Thereafter a similar calculation to the $\Delta$ and $\mu$-proportional terms, we find
\begin{eqnarray}
H&=&\sum_{j=1}^{N-1}\frac{t i}{2}(\eta_{j+1}\gamma_j-\gamma_{j+1}\eta_j)-\frac{\Delta i}{4}(\eta_{j+1}\gamma_{j}+\gamma_{j+1}\eta_j)+\nonumber \\
&&\frac{\mu i}{2}\sum_{j=1}^{N}\gamma_j \eta_j. \label{KiMajoop}
\end{eqnarray}  

For the sake of simplicity, we shall consider $\mu=0$ and $t=\Delta/2$ in Eq.~(\ref{KiMajoop}). Therefore,
\begin{equation}
H=-\frac{\Delta i}{2}\sum_{j=1}^{N-1}\gamma_{j+1}\eta_j. \label{ntpki}
\end{equation}

Remarkably, the competition between the hopping parameter $t$ and the coupling $\Delta$ has eliminated two boundary-Majorana operators. For instance, let us assume $N=3$, it follows from Eq.~(\ref{ntpki}) that
\begin{equation}
H=-\frac{\Delta i}{2}(\gamma_2 \eta_1+\gamma_3 \eta_2),
\end{equation}  
where the boundary-Majorana operators $\gamma_1$ and $\eta_3$ have been eliminated. They are, therefore, zero-energy modes at the boundaries of the quantum wire, implying a nontrivial topological phase in the Kitaev model. 

For a more general result, we may apply the Majorana number, defined in Eq.~(\ref{Majonumber}), for the Kitaev Hamiltonian in Eq.~(\ref{Kibloch}).  After simple calculations,  we find that $M=\sgn(\mu-2t)$. Therefore, for $2t>\mu$, the system has Majorana fermions.


The Kitaev model is considered the theoretical realization of Majorana fermions in condensed matter physics. The realization of Majorana quasiparticle may have a important application to a new area called topological quantum computation.

\section{A.~VI-Topological phases in the Dirac and the BHZ theories.}
\label{cap57} 

In this Section, we calculate the Chern number for both Dirac and BHZ models. These are common examples of theories with topological properties in (2+1) dimensions. Besides that, the Dirac theory applies in the low-energy description of graphene and BHZ model applies for the description of the quantum spin Hall effect.

\subsection{Dirac Theory and Parity Anomaly}

The low-energy theory for describing  the p-electrons, in the honeycomb lattice of graphene, is the massless Dirac theory. Here, we explore the the topological properties of the massive Dirac theory, which may be understood as graphene with a gap generation.

The Dirac Hamiltonian is
\begin{equation}
H=\vec{\sigma}.\textbf{h}(\textbf{k}), \, \textbf{h}=(k_x,k_y,m) \label{diraccont}
\end{equation}
with ``Chern number'' given by Eq.~(\ref{chn}). Thereafter a simple algebra, we obtain
\begin{equation}
n=\frac{1}{4\pi}\int_0^{\infty}\int_0^{2\pi}k dk d\phi \frac{m}{(k^2+m^2)^{3/2}}, \label{qmchn}
\end{equation}
where we must integrate out over the whole momentum space. Therefore, the ``Chern number'' for the massive Dirac theory in two-dimensions is
\begin{equation}
n=\frac{\sgn(m)}{2}.  \label{chndirac2d}
\end{equation}

Because the Hall conductivity is, essentially, given by the Chern number of the filled energy bands, accordingly with Eq.~(\ref{chnhall}), one would conclude that for a single Dirac Hamiltonian, there exist a quantum Hall effect with conductivity of $\sgn(m)e^2/2h$. Furthermore, one may perform the massless limit, $m\rightarrow 0$, in Eq.~(\ref{diraccont}) and Eq.~(\ref{chndirac2d}), finding that, even for a massless theory, the Hall conductivity is nonzero. Because the massless Dirac theory is invariant by time-reversal symmetry, but the Hall current breaks this symmetry, hence, such result is a parity anomaly. The very same result may be obtained from the Kubo formula, where the Hall conductivity is given by the coefficient of the Chern-Simons term of the vacuum polarization tensor. Indeed, using dimensional regularization, it is well known that this coefficient is $\sgn(m)/2$. Next, we have to consider that graphene has two valleys K and K', which are connected by time-reversal symmetry. This symmetry changes the sign of the masses, then the total Hall conductivity \textit{vanishes}. 

Although the result in Eq.~(\ref{chndirac2d}) is correct, it is not a proper Chern number, which is the reason why we have used quotation marks above Eq.~(\ref{qmchn}). Indeed, we have shown in Section A.~II that the Chern number is mapping from the Brillouin zone to the sphere in the h-space. Therefore, we need a closed surface (The Brillouin zone) that is absent in the Dirac theory. Physically, it implies a half-quantized Hall conductivity either $+e^2/2h$ or $-e^2/2h$, which is a puzzling result. 

The correct procedure to calculate the Chern number is to perform the calculations in the lattice theory, after that to apply the desired limits. We would like to emphasize that this \textit{has been done} in Ref.~\cite{Luscher} for the Dirac theory. Indeed, the authors have shown that, by using the lattice parameter $a$ as a natural regulator, one obtains a quantized Hall conductivity, using the Kubo formula with the vacuum polarization tensor that they have calculated. Therefore, the quantization of the Hall conductivity is related to the lattice regularization of the parity anomaly. In the massless case $m=0$, it yields a spontaneous quantum valley Hall effect \cite{prx}. In this case, this valley conductivity is quantized exact as the Hall conductivity for graphene $\sigma^V_{xy}=4(n+1/2)e^2/h$ and Landau-like energy levels are dynamically generated due to the electromagnetic interaction.  The result of Ref.~\cite{Luscher} solves this long misunderstand between anomalous quantum Hall effect and parity anomaly. Indeed, using their result, a well-quantized Hall conductivity emerges from the Kubo formula \cite{prx}. 

\subsection{BHZ Model}

Since the Dirac theory has nontrivial topology, we hope that its lattice generalization also admits this property. Here, we use the lattice generalization of the Dirac theory that yields  Ref.\cite{Bernevig}
\begin{equation}
k_x, k_y\rightarrow A\sin k_x, A\sin k_y, \, m\rightarrow m-2B(2-\cos k_x-\cos k_y).
\end{equation}
Therefore, the Dirac-lattice Hamiltonian is
\begin{equation}
H(\textbf{k})=\vec{\sigma}. \textbf{D}(\textbf{k}), \label{diraclattice}
\end{equation}
with
\begin{equation}
\textbf{D}(\textbf{k})=\textbf{(}A\sin k_x,A\sin k_y,m-2B(2-\cos k_x-\cos k_y)\textbf{)}.  \label{DBHZ}
\end{equation}

The parameters $A$ and $B$ are constraints for comparison with the BHZ model \cite{BHZ,BHZexp}. In the lowest order Eq.~(\ref{diraclattice}) yields the usual Dirac theory. Finally, we suppose two independent copies of the lattice Dirac theory, the lattice Dirac Hamiltonian $H(\textbf{k})$ and its time reversal part $H^*(-\textbf{k})$. This is the BHZ model, which the Hamiltonian reads
\begin{equation}
H_{\rm{BHZ}}=\left(\begin{matrix} H(\textbf{k})& 0 \\ 0& H^*(-\textbf{k}) \end{matrix}\right), \label{BHZ}
\end{equation}
where $H(\textbf{k})$ is given by Eq.~(\ref{diraclattice}).

Next, we calculate the Chern number for the BHZ model. Nevertheless, we use Eq.~(\ref{geochn}) that provides the Chern number dependent on a sum over high-symmetry points instead of the usual expression in Eq.~(\ref{chn}). Before, we apply this result to BHZ model, we briefly discuss how to obtain the Chern number for the continuum Dirac theory, using this equation. The first step is to identify the high-symmetry points $D_i$. In order to do so, we have to choose points in which only one of the axis $h_1$, $h_2$, and $h_3$ are not trivial. 

For the continuum Dirac theory, it is easy to check that $D_i=(k_x,k_y)=(0,0)$, because $h_1(D_i)=h_2(D_i)=0$ and $h_3(D_i)=m$. The second step is to calculate the Jacobian in Eq.~(\ref{jac}) at all $D_i$ points, which means we only need the component $i=3$ of the Jacobian, because $h_3$ is the relevant axis. Using Eq.~(\ref{diraccont}), we have $J_3=\hat {h}_3$, thereby $\sgn(J_3)=1$. Using Eq.~(\ref{geochn}), we obtain the Chern number $n=\sgn(m)e^2/2h$ for the continuum Dirac theory, in agreement with Eq.~(\ref{chndirac2d}).

For the BHZ model, the high symmetry or Dirac points $D_i$ are given by $D_i=\{(k_x,k_y)\}=\{(0,0),(0,\pi),(\pi,0),(\pi,\pi)\}$, where $h_1(D_i)=h_2(D_i)=0$ and $h_3(D_i)=\{m,m-4B,m-4B,m-8B\}$. Using Eq.~(\ref{BHZ}) and Eq.~(\ref{DBHZ}), we obtain that the sign of Jacobian $J_i$ for $i=3$ is $\sgn(J_3)=\sgn[\cos (k_x) \cos(k_y)]$. Therefore, we have the so-called chirality of the model, given by $\sgn[J_3(D_i)]=\{+,-,-,+\}$. Using Eq.~(\ref{geochn}), we obtain the Cher number for the BHZ model
\begin{equation}
n=\frac{1}{2}[\sgn(m)+\sgn(m-8B)-2 \,\sgn(m-4B)].
\end{equation}
Therefore, $n=1$ for $m \in (0,4B)$ and $n=-1$ for $m \in (4B,8B)$, otherwise $n=0$ and the model is trivial. This result is for one copy of the BHZ model, the total Chern number should be multiplied for $2$, thereby, we may obtain $n=2,-2$.

\section{Part B: Electronic Interactions}

In this second part of the paper, we discuss the role of electronic interactions for describing topological phases. In order to do so, we shall use the PQED approach for graphene at low energies.

We have shown in Sec.~A.IV that the transversal conductivity is given by a sum over all the energy levels below the Fermi level. Hence, a discrete spectrum generates a Hall conductivity. For instance, graphene with a applied magnetic field $B$ exhibits quantized energy levels, given by $E_n=\hbar \omega_D \sqrt{n}$ with $\omega_D=v_F\sqrt{e B/\hbar c}$. These are known as relativistic Landau levels. This quantization implies a Hall conductivity given by $\sigma^H_{xy}=(n+1/2)e^2/h$, which may be experimentally observed at room temperatures \cite{RTQHE}. In this case, the external magnetic field changes the electronic spectrum from the Dirac cones to gapped energy levels. Furthermore, there exist an explicit breakdown of time-reversal symmetry due to $B$, in agreement with Haldane's condition for existence of quantum Hall effect \cite{Bernevig}.  

Could one generate quantized energy levels in the absence of $B$? Because electrons are charged, the natural interaction among them is the electromagnetic interaction. Using the PQED approach for graphene, we show that the electronic spectrum is renormalized at large enough coupling constant $\alpha$, i.e, if $\alpha>\alpha_c$ with $\alpha_c\approx 1.02$. The dynamical mass generates a set of quantized energy levels $m^*_n$, yielding a quantized valley Hall conductivity $\sigma^V_{xy}=(n+1/2)e^2/h$ \cite{prx}. In order to preserve the time-reversal symmetry (because $B=0$), the Hall current vanishes. Hence, we conclude that the two valleys of graphene are counter-propagating at the edges. The quantum valley Hall phase is a topological state of matter, very similar to the quantum spin Hall.

\section{B.~I- Electronic Interactions and Renormalization of the Electron Energy Spectrum}

We have shown, in Section A.~IV, that the sum of the Berry phase over the set $l_0$ of filled energy bands gives the Hall conductivity. Hence, the knowledge of the electron-energy spectrum is essential. It turns out that, within a interacting picture, this spectrum may drastically change, which is the case in the strong correlated limit. Here, we shall discuss only how the electronic interactions changes the electronic spectrum.

The Lagrangian model for describing electronic interactions in (2+1) dimensions is given by
\begin{equation}
{\cal L}= \frac{1}{2} F_{\mu \nu} (-\Box)^{-1/2} F^{\mu\nu}+{\cal L}_M[\psi]+e j^{\mu}A_{\mu}\, ,
\label{action}
\end{equation}
where $F_{\mu\nu}=\partial_\mu A_\nu-\partial_\nu A_\mu$ is the usual field intensity tensor of the U(1) gauge field $A_\mu$, which intermediates the electromagnetic interaction in 2+1 dimensions (pseudo electromagnetic field). ${\cal L}_M[\psi]=\bar\psi (i\partial\!\!\!/-m_0) \psi$ is Dirac Lagrangian, $j_\mu=\bar\psi\gamma^\mu\psi$ is the matter current, $e^2=4\pi\alpha$ is the electric charge, and $\alpha$ the fine-structure constant. For graphene, since the Fermi velocity is $c/300$, we find $\alpha_g\approx 300/137 \approx 2.2$. Although we specify the matter field, the gauge field is obviously independent on the fermion term. The nonlocal operator reads
\begin{equation}
(-\Box)^{-1/2}\equiv \int \frac{d^3k}{(2\pi)^3}\exp(-i k x)\frac{1}{\sqrt{k^2}}. \label{nlop}
\end{equation} 

The Feynman rules are
\begin{equation}
G_{0,\mu\nu}=\frac{1}{2\sqrt{p^2}}\left(\delta_{\mu\nu}-\frac{p_\mu p_\nu}{p^2}\right) \label{gprop}
\end{equation}
for the gauge-field propagator,
\begin{equation}
S_{0F}=\frac{1}{\gamma^\mu p_{\mu}}=\frac{\gamma^\mu p_{\mu}}{p^2_0-\textbf{p}^2}, \label{mprop}
\end{equation}
for the electron propagator, and $e\gamma^\mu$ for the vertex interaction, where $\gamma^\mu$ are the Dirac matrices. From Eq.~(\ref{mprop}), we find the pole of the fermion propagator $p_0\equiv E(\textbf{p})=\pm \sqrt{c^2 \textbf{p}^2+m^2_0}$. For comparison with graphene, we set $m_0=0$ and $c\rightarrow v_F$, yielding $E(\textbf{p})=\pm v_F |\textbf{p}|$, the two band- and massless- spectrum of electrons in graphene. Note that the main difference, in comparison with QED, is the denominator $\sqrt{p^2}$ instead of $p^2$ as in the Maxwell theory. This has been derived in Ref.~\cite{marino}, assuming only that the matter current is confined in the plane, which holds for all 2D materials. 

For the sake of physical interpretation, we may calculate the static limit of the model in Eq.~(\ref{action}). Let us define the functional ${\cal Z}$ of the theory, given by
\begin{equation}
{\cal Z}=\int DA_\mu D\bar\psi D\psi \exp\left[\int d^3x {\cal L}(A_\mu,\bar\psi,\psi)\right] \label{zfunctional}.
\end{equation}
Next, we integrate out $A_\mu$ in Eq.~(\ref{zfunctional}), yielding
\begin{equation}
{\cal Z}=\int D\bar\psi D\psi \exp\left[\int d^3x {\cal L}_{{\rm eff}}(\bar\psi,\psi)\right],
\end{equation}
where the effective action ${\cal L}_{{\rm eff}}(\bar\psi,\psi)$ is given by
\begin{equation}
{\cal L}_{{\rm eff}}(\psi)={\cal L}_M[\psi]+e^2j^\mu G_{0,\mu\nu} j^\nu.
\end{equation}
In the static limit, we must consider $j^\mu(t,\textbf{r})\rightarrow j^\mu(\textbf{r})=(j_0,\textbf{j}=0)$, because there is no current. This also implies $p_0=0$ for the gauge-field propagator. Hence, using Eq.~(\ref{gprop}), we obtain that the static potential $V(r)$ is only the Fourier transform of $1/\sqrt{\textbf{p}^2}$, namely,
\begin{equation}
V(r)=e^2\int \frac{d^2\textbf{p}}{(2\pi)^2}e^{i\textbf{p}.\textbf{r}} G_{0,\mu\nu}(p_0=0,\textbf{p})\delta_{\mu 0} \delta_{\nu 0},
\end{equation}
which yields
\begin{equation}
V(r)=\frac{e^2}{4\pi r},
\end{equation}
the physical Coulomb potential, generated by Eq.~(\ref{action}). On the other hand, by using the Maxwell propagator in (2+1) dimensions, one may obtain a (unphysical) logarithmic potential $\ln (e^2 r)$ \cite{Appel}. Next, we consider the dynamical limit for calculating interaction effects. In particular, we would like to investigate the possibility of generating a dynamical energy gap for the electrons.
 
We have to calculate the renormalized electron-energy spectrum. This may be obtained with the Schwinger-Dyson equation for the electron propagator. The Schwinger-Dyson equation for the full electron propagator reads
\begin{equation}
S_F^{-1}(p)=S_{0F}^{-1}(p)-\Xi(p),  \label{fullpropfer2cap5}
\end{equation}
where $S_F(p)$ is the full electron propagator, respectively. The term ``full'' means it has the interaction corrections. The electron self-energy $\Xi(p)$ is given by
\begin{eqnarray}\label{fermioncap5}
\Xi (p)=e^2\int\frac{d^3k}{(2\pi)^3} \gamma^{\mu}S_F(k) \gamma^{\nu}\,G_{\mu\nu}(p-k),
\end{eqnarray}
and $G_{\mu\nu}$ is the full gauge-field propagator. This is obtained from $G^{-1}_{\mu\nu}=G^{-1}_{0,\mu\nu}-\Pi^{\mu\nu}$, where $\Pi^{\mu\nu}$ is the vacuum polarization tensor. $\Pi^{\mu\nu}$ has been calculated in Ref.~\cite{Luscher} in one loop. Therefore, we must bear in mind that we know the full gauge-field propagator. Next, let us find the full electron propagator. 

Before we specify any kind of approximation, let us calculate the pole of the full-electron propagator. We perform a Taylor expansion in the electron self-energy in Eq.~(\ref{fermioncap5}) around $m^*$, yielding 
\begin{eqnarray}
\Xi(p)=\Xi(p=m^*)+(\gamma^\mu p_\mu-m_R)\frac{\partial\,\Xi(p)}{\partial p}\huge|_{p=m^*}+...\end{eqnarray}

Using
\begin{equation}
\delta m=m^* -m_0, \label{cond1}
\end{equation}
where $\delta m\equiv \Xi(p=m^*)$, we may write the full fermion propagator, in Eq.~(\ref{fullpropfer2cap5}), as
\begin{eqnarray}
S_F(p)&=&\frac{1}{\gamma^\mu p_\mu-m_0-\Xi(p)}\nonumber\\
&=&\frac{1}{(\gamma^\mu p_\mu-m^*)(1-\frac{\partial\,\Xi(p)}{\partial p}\huge|_{p=m^*}+...)}
\nonumber\\
&=&\frac{\gamma^\mu p_\mu+m^*}{(p^2 - m_*^2)(1-\frac{\partial\,\Xi(p)}{\partial p}\huge|_{p=m^*}+...)}.
\label{sfsup}
\end{eqnarray}
We find that  $m^*$ is the pole of the full physical electron propagator at zero momentum, being therefore
the desired physical mass, i.e, the renormalized electronic spectrum is $E(\textbf{p})=\pm\sqrt{\textbf{p}^2+m_*^2}$. This result requires no approximation. 

Within the perturbation expansion in $\alpha$, Eq.~(\ref{cond1}) yields $m^*=\Xi(m_0)$. In this case, the most general solution for the electron self-energy reads $\Xi(p)=A(p)\gamma^\mu p_\mu$, where $A(p)$ is an arbitrary function. Hence, in the massless limit $m_0\rightarrow 0$, we find $m^*=0$, which is not surprisingly. Indeed, it is well known that there is no dynamical mass generation in perturbation theory \cite{Appel}. 

For large coupling constant, the previous result does not apply. In this case, the most general solution reads $\Xi(p)=A(p)\gamma^\mu p_\mu-\Sigma(p)$, therefore, the mass is $m^*=A(m^*)m^*-\Sigma(p^2=m^2_*)$, where $\Sigma(p)$ is called the mass function \cite{Appel}. For a more details about the calculation of $\Sigma(p)$ for PQED with massless Dirac particles, see Ref.~\cite{prx}.

Using well-known approximations for U(1) Gauge theories, the large coupling limit has been calculated in Ref.~\cite{prx} for the action in Eq.~(\ref{action}). The authors have shown that a set of Landau-like energy levels are generated for $\alpha>\alpha_c\approx 1.02$. These energy levels are given by
\begin{equation}
m^*_n=\pm \Lambda \exp\left(-\frac{Z_n}{2\sqrt{\alpha/\alpha_c-1}}\right), \label{massesd}
\end{equation} 
where $\Lambda=\hbar v_F/a\approx 3$ eV is the lattice cutoff, $v_F\approx c/300$ is the Fermi velocity, $a\approx 10^{-10}$ m is the lattice parameter, and $Z_n=C_0 (2n+1)\pi/2$ with $C_0>0$, being some known  constant. 

\begin{figure}[htb]
\label{massesgra}
\centering
\includegraphics[scale=0.42]{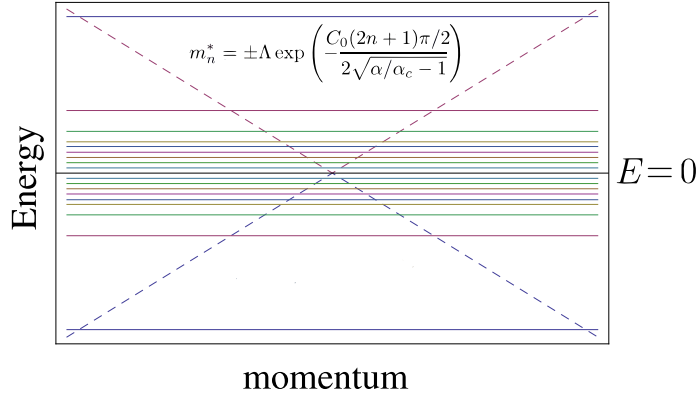}
\caption{Schematic representation of the dynamically generated masses in graphene, given by Eq.~(\ref{massesd}). The solutions are symmetric around the Dirac point $E=0$ and the largest energy gap is $|m^*_0|$. The two valleys are connected by time-reversal symmetry, therefore, they have opposite masses. For large $n$, the energy levels are close to the Dirac point. This figure resembles the relativistic Landau levels for which the spacing between the energy levels is proportional to $\sqrt{n}$, accumulating for large $n$. }
\end{figure}

\begin{figure}[htb]
\label{QVHE}
\centering
\includegraphics[scale=0.35]{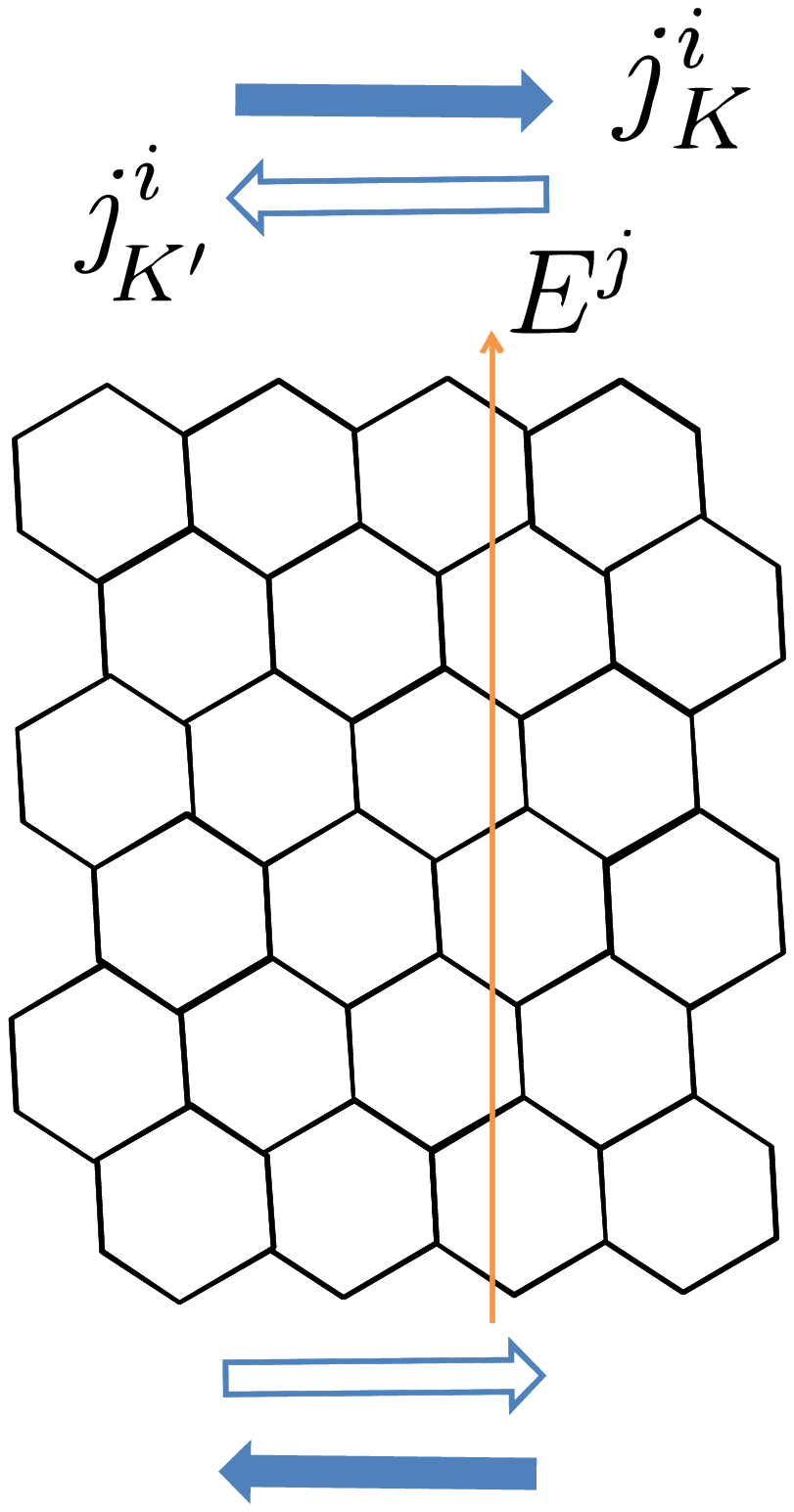}
\caption{Schematic representation  of the two counter-propagating valley currents in graphene. The difference of the valley current generates the quantum valley Hall effect, with the conductivity given in Eq.~(\ref{svxy}). This valley current propagates on the edge of the graphene sheet. Since the effect is generated by the masses in Eq.~(\ref{massesd}), hence  the presence of an external magnetic field is not required.} \label{fig5}
\end{figure}

These energy levels collapse around $E=0$ for $n\rightarrow \infty$, i.e, $\Delta m^*_n=m^*_{n+1}-m^*_n \rightarrow 0$ for large $n$. On the other hand, the relativistic Landau levels are proportional to $\sqrt{n}$. Hence, $\Delta E_n\propto \sqrt{n+1}-\sqrt{n}\rightarrow 0$. It shows that both results have the same kind of gapped structure. In principle, by fine tuning some parameters, one would obtain $\Delta m^*_n=\Delta E_n$ for some $n$. In other words, there exist a continuous transform that changes the energy levels $\Delta m^*_n$ to $\Delta E_n$ without closing the energy gaps. Hence, in the light of the quantum adiabatic theorem, the Berry phase is the same and also the topological properties. Using the Kubo formula, it has been shown that a set of quantized valley Hall conductivity 
\begin{equation}
\sigma^V_{xy}=4\left(n+\frac{1}{2}\right)\frac{e^2}{h} \label{svxy}
\end{equation}
emerges due to this set of Landau-like generated masses, see Fig.~\ref{fig5}. The total Hall conductivity vanishes because the two valleys are connected by time-reversal symmetry. Nevertheless, by using a valley filter, which turn off one of the valley currents, it is possible to observe this nice result, which as far as we know, is the only one connecting topological phases with dynamical mass generation.

\section{B.~II DC Conductivity of PQED}

We use PQED in Eq.~(\ref{action}) in order to obtain both the longitudinal and the Hall conductivity. Using the minimal principle for $A_\mu$ in Eq.~(\ref{action}), we have
\begin{equation}
\langle \frac{\delta(1/2 F_{\mu \nu} (-\Box)^{-1/2} F^{\mu\nu} )}{\delta A_\mu}\rangle=-\langle j^\mu \rangle. \label{mpA}
\end{equation}

Integrating the fermion field in Eq.~(\ref{action}), it is straightforward that the effective action for $A_\mu$ reads 
\begin{equation}
{\cal L}_{{\rm eff.}}=\frac{1}{2} F_{\mu \nu} (-\Box)^{-1/2} F^{\mu\nu}+\frac{A_\mu \Pi^{\mu\nu}A_\nu}{2}, \label{actioneff}
\end{equation}
where $\Pi^{\mu\nu}$ is the vacuum polarization tensor. In general grounds, we may write this tensor as
\begin{equation}
\Pi_{\mu\nu}=A(p)P_{\mu\nu}+B(p)\epsilon_{\mu\nu\alpha}p^\alpha.
\end{equation}

Using Eq.~(\ref{mpA}) and the minimal principle of the action in Eq.~(\ref{actioneff}), we find
\begin{equation}
\langle j_\mu \rangle=\Pi_{\mu\nu} A^\nu. \label{curm1}
\end{equation}

Using the optical limit, i.e, $\textbf{p}\rightarrow 0$ in Eq.~(\ref{curm1}) and $E_k(\omega)=\omega A_k(\omega)$, we find that the electric current reads
\begin{equation}
\langle j_i \rangle=\left[\frac{A(\omega)}{\omega}\delta_{ik}+B(\omega)\epsilon_{ik0}\right]E^k,
\end{equation}
where $\epsilon_{ik0}$ is the anti-symmetric tensor and $E_k$ is an external field. The optical limit is obtained with $\textbf{p}\rightarrow 0$ and, after this, the dc limit is performed with $\omega\rightarrow 0$. Therefore, the longitudinal conductivity (using the vacuum polarization tensor in two-loop) reads \cite{prx} 
\begin{equation}
\sigma^{xx}= \left(\frac{\pi}{2} \frac{e^2}{h}\right)\left[1+\left(\frac{92 - 9 \pi^2}{18\pi} \right)\,\alpha_g
+{\cal O}(e^4)\right].  \label{imp11}
\end{equation}

\begin{figure}[htb]
\label{cond}
\centering
\includegraphics[scale=0.30]{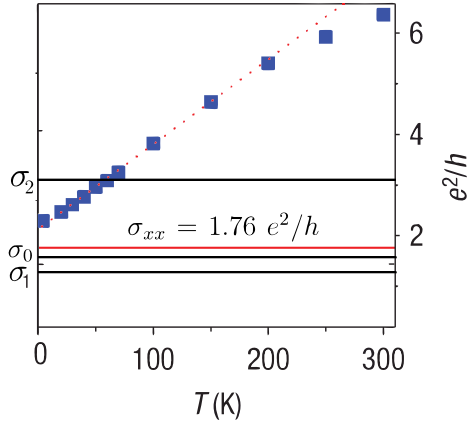}
\caption{The experimental measurement of the longitudinal conductivity $\sigma_{xx}$ in units of $e^2/h$ as a function of temperature $T$ in Kelvin. The filled squares are experimental points, and the dashed red line is a linear fit to these data. The solid red line is our theoretical result with quantum correction to the zero-temperature result. The solid black lines are other theoretical results which have been discussed in Ref.~\cite{Ziegler}. These results are $\sigma_2=\pi e^2/h \approx 3.14 e^2/h$, $\sigma_1=4e^2/\pi h\approx 1.27 e^2/h$, and $\sigma_0=\pi e^2/2 h\approx 1.57 e^2/h$. Note that we have used lines to represent these theoretical results in order to facilitate the visualization; however, these results only hold in the zero-temperature limit. The experimental data have been extracted from Ref.~\cite{Andrei2}.}\label{fig6}
\end{figure}

Eq.~(\ref{imp11}) yields the correction provided by PQED to the non-interacting value $\sigma_0 = \pi e^2 / 2h$, obtained in Ref.~\cite{theorynonint}. This result, namely, $\sigma_{xx} = 1.76$ $e^2/h$ (for $\alpha_g\approx 2.2$ for graphene) is the closest theoretical result in comparison with the experimental data for the conductivity extrapolated to zero temperature: $\sigma_{xx} = 2.16$ $e^2/h$ obtained in Ref.~\cite{Andrei2}; see Fig.~\ref{fig6}. In this figure, we have compared Eq.~(\ref{imp11}) with other theoretical predictions for $\sigma_{xx}$ in graphene Ref.~\cite{Ziegler}. These results differ from each other because the longitudinal conductivity is sensitive to the order of the limits, related to: temperature $T$, disorder $\eta$, and frequency $\omega$. The noninteracting value $\sigma_{xx}=\sigma_0$, which was predicted in Ref.~\cite{theorynonint}, reads
\begin{equation}
\sigma_{0}=\lim_{\omega\rightarrow 0} \lim_{\eta\rightarrow 0} \sigma_{xx}(T=0,\eta,\omega)=\frac{\pi}{2} \frac{e^2}{h}.
\end{equation}
Note that this result is obtained if we perform first both the zero-temperature and zero-disorder limits, thereafter these limits, we perform the zero-frequency limit. Conversely, if we consider first the zero-temperature limit with finite disorder equal to the frequency $\omega$, we have  
\begin{equation}
\sigma_{2}=\lim_{\omega\rightarrow \eta} \sigma_{xx}(T=0,\eta,\omega)=\pi \frac{e^2}{h}.
\end{equation}
In this case, the disorder effect has been considered in the final result. In the last example, if we first consider the zero-frequency limit and then the zero-disorder limit, it yields 
\begin{equation}
\sigma_{1}=\lim_{\eta\rightarrow 0} \lim_{\omega\rightarrow 0} \sigma_{xx}(T=0,\eta,\omega)=\frac{4}{\pi} \frac{e^2}{h}.
\end{equation}

\section{ \textbf{II.\, Summary and Outlook}}

We have provided a brief introduction to the main concepts of the topological insulator theory, such as: Berry phase, Chern number, and quantum adiabatic theorem. For one-dimensional system, we also reviewed the definition of Majorana number and applied it to the famous Kitaev Chain model. 

For two-dimensional materials, we have derived that the Hall conductivity is given by the sum of the Berry phase of all filled energy bands. Because the electronic interaction may dynamically change the electron energy spectrum, hence the number of filled energy bands, we have concluded that electronic interactions are relevant for describing topological phases at large coupling regime. 

The topological insulator theory has been attracted great attention in literature. Nevertheless, electronic interactions, particularly with a dynamical approach as in PQED, has been less discussed. There are some aspects that support the dynamical approach: (a) There is no reason to believe that the dynamical regime lacks of the physical results in the static limit; (b) It has been experimentally shown that in the low doping limit, the Fermi velocity for graphene increases \cite{Elias}, yielding a better regime for comparison with our approach; (c) Without a complete vertex interaction $j^\mu A_\mu$, some parity anomalies may not be obtained. Indeed, the Chern-Simons term proportional to $\epsilon^{\mu\nu\alpha}A_\mu \partial_\nu A_\alpha$ vanishes for $\mu=\alpha=0$, leading to a zero valley Hall conductivity; (d) Only this dynamical regime may be applied to compare with recent proposals for realization of gauge theories in (2+1) dimensions using ultracold atoms \cite{coldatoms}. A possible generalization of the results in Ref.~\cite{prx} would be to investigate the PQED on a lattice and electronic interactions in other 2D materials.

\section{\textbf{acknowledgments}}

This work was partly supported by: Ministry of Science, Technology and Innovation of Brazil (MCTI-Brazil); Ministry of Education and Culture of Brazil (MEC-Brazil); The program ``Science without Borders'' of National Council for Scientific and Technological Development (CNPQ-Brazil).  I am grateful to E. C. Marino, V. S. Alves, C. Morais Smith, T. Macri, R. G. Pereira, and L. Fritz for very interesting and stimulating discussions.

\end{document}